\documentclass[aps,prx,twocolumn,superscriptaddress,amsmath,amssymb,longbibliography]{revtex4-2}

\usepackage{graphicx}
\usepackage{dcolumn}
\usepackage{bm}
\usepackage{physics}
\usepackage{xcolor}
\usepackage{amsthm}
\usepackage{hyperref}

\newtheorem{theorem}{Theorem}

\newcommand{\Dd}{\mathcal D}
\newcommand{\cU}{\mathcal U}

\begin{document}
\title{Testing non-classical gravitational phase dynamics under strong dephasing}
\author{Tianfeng Feng}
\email{fengtf@hku.hk}
\affiliation{QICI Quantum Information and Computation Initiative, School of Computing and Data Science, The University of Hong Kong, Pokfulam Road, Hong Kong SAR, China}
\author{Vlatko Vedral}
\affiliation{Department of Physics, University of Oxford, Parks Road, Oxford OX1 3PU, United Kingdom}
\date{\today}

\begin{abstract}
Whether gravity must be quantum is often approached by asking whether two masses can become entangled through their mutual gravitational interaction. In practice, the weak interaction competes with environmental decoherence, so the final state may contain no entanglement even when the underlying gravitational phase remains coherent. We investigate what can still be learned from the dynamics in a dephased Bose--Marletto--Vedral setup. The dominant Newtonian effect for well-separated paths is an accumulated phase rather than transfer between interferometer arms, a regime motivated by matter-wave gravitational phase experiments and quantified here by wavepacket-overlap and deflection estimates. Keeping independently characterized local which-path dephasing fixed, we prove an exact finite-time distinction between the gravitational conditional-phase evolution and every passive stochastic local-phase model (induced by a classical mediator), including regimes where the chosen input remains separable at all times. The distinction survives, with a modified sensitivity, when the comparison is enlarged to population-preserving classical measurement and feedback through a diagonal PPT relaxation. These results identify an operational test of gravitational phase dynamics in the strong-dephasing regimes, conditional on explicit and physically motivated classical mediator models.
\end{abstract}
\maketitle

The question of whether gravity must obey quantum principles predates the modern search for experimental quantum-gravity effects. At the 1957 Chapel Hill conference, Feynman asked what would happen if a sufficiently massive body were placed in a coherent superposition of two positions, each associated with a different gravitational field. His thought experiment made a point that remains central today: a gravitational field linked to a quantum superposition confronts us with a question about the coherence of the interaction itself, not merely with the size of quantum corrections to a classical trajectory \cite{feynman1957,pageGeilker1981,marlettoVedral2025RMP}. The Page--Geilker experiment later challenged the simplest expectation-value version of semiclassical gravity, while leaving room for more sophisticated alternatives \cite{pageGeilker1981}. Neither observation establishes that a quantized gravitational field has already been seen, resulting in a long-standing debate over whether gravity should be quantized \cite{dewitt2003global,eppley1977necessity,peres2001hybrid,marletto2017we,feng2026conservationlawsnonclassicalitygravity}.
The physical challenge is to design a laboratory test in which competing descriptions make different operational predictions.

Matter-wave interferometry provides an unusually direct route to that challenge because phases can be easier to observe than gravitational displacements. In a gravitational potential, each quantum path acquires an action phase, and interferometric recombination converts a relative phase into a detectable population modulation. The Colella--Overhauser--Werner (COW) experiment measured the Earth's gravitationally induced phase of neutron matter waves, and subsequent neutron interferometry refined the same basic approach \cite{COWPhysRevLett.33.1237,cow1975,abeleleeb2012}. More recently, atom interferometry observed a gravitational Aharonov--Bohm effect due to an external source mass, using a configuration designed to separate the accumulated phase from force-induced path motion \cite{overstreet2022}. The contrast is instructive: a scalar Newtonian potential can be a very effective phase shifter even when the associated deflection is extremely small. The coherent wave itself is the sensitive gravimeter and a large transfer of population between spatially separate arms is not required.

The reach of this phase-first approach is no longer confined to microscopic interferometers. Experiments have demonstrated interference of increasingly complex molecules, and precision measurements have directly resolved the gravitational coupling of small mechanical masses \cite{fein2019,westphal2021}. The ability to preserve coherence in heavy objects is nevertheless limited by collisions, thermal photon emission and other sources of which-path information, as established in matter-wave decoherence experiments and microscopic treatments \cite{hornberger2003,hackermueller2004,hornbergerSipe2003,hornbergerSipeArndt2004}. These advances motivate a careful separation between what gravity does coherently and what the external environment erases.

The Bose--Marletto--Vedral (BMV) proposals bring the question of gravitational coherence into a genuinely two-body setting \cite{bose_spin_2017,marletto_gravitationally_2017}. Earlier proposals had already explored coherence in gravitational fields and Bose--Einstein condensates \cite{lindnerPeres2005}, but the two-mass proposal targets a particularly transparent entangling conditional phase. Instead of treating one gravitational source as a prescribed classical object, both masses are coherently split into two paths. The potential energy depends on which pair of paths is occupied, giving four branch-dependent phases. An overall phase has no observable consequence, and the phases depending on only one mass can be absorbed into local reference frames. What remains is a conditional two-body phase. In the standard quantum Newtonian model, it is generated by a $Z_AZ_B$ interaction and can entangle initially separable path superpositions \cite{bose_spin_2017,marletto_gravitationally_2017}. Under a locality assumption that limits a classical mediator to local operations and classical communication, entanglement generated solely by the interaction would witness a non-classical mediation resource \cite{CMVVPhysRevD.102.086012,marletto_gravitationally_2017,marlettoVedral2025RMP, krisnanda2017,kafri2013noiseinequalityclassicalforces,galley2022,marshman2020}. The precise meaning of that resource and its dependence on the assumed mediating dynamics continue to be examined \cite{danielson2022,christodoulou2023,christodoulou2023Degrees,martinmartinez2023,carney2022,fragkos2022,r8ry-sp35,altamirano2018gravity}.

Realizing the ideal BMV signature is difficult because the interaction is extraordinarily weak. Proposed routes include spin-controlled matter-wave superpositions, shielding of electromagnetic backgrounds, dynamical decoupling and enhanced gravitational coupling \cite{boseRoadmap2025,vandeKamp2020,pedernales2020,pedernales2022,cosco2021,weiss2021,fengVedral2022,wv19-5t6p}. Independent proposals study alternative optomechanical or oscillator platforms, quantum optical correlations, non-Gaussian signatures and refined witness protocols \cite{krisnanda2020,miao2020,chevalier2020,howl2021,carlesso2019,haine2021,plato2023enhanced, sidajaya2022,noman2026operational,wang2026recoverabilitygravitationalentanglementdecoherence}. Most of these approaches differ in implementation but share a severe constraint: the accumulated conditional phase must compete with dephasing throughout the interaction time. Consequently a nonzero conditional gravitational interaction need not produce an entangled output. For the coherent BMV probe, sufficiently strong local which-path noise can prevent entanglement from appearing at any interrogation time, even though a finite conditional phase continues to be accumulated.

This is where a distinction between states and dynamics becomes decisive. The entanglement of an output state records whether the chosen initial state remains nonseparable at the point of observation. It does not by itself determine whether the interaction responsible for that output admits a specified classical implementation. Once the final state is separable, a machine permitted to prepare that state directly can reproduce it. But state preparation is not the same task as reproducing a physical gravitational mechanism in the presence of a separately characterized environment. The meaningful question is whether the conditional phase evolution, with ordinary dephasing held fixed, can be replaced by the dynamics of a classical mediator.

\begin{figure*}[t]
\centering
\includegraphics[width=0.93\textwidth]{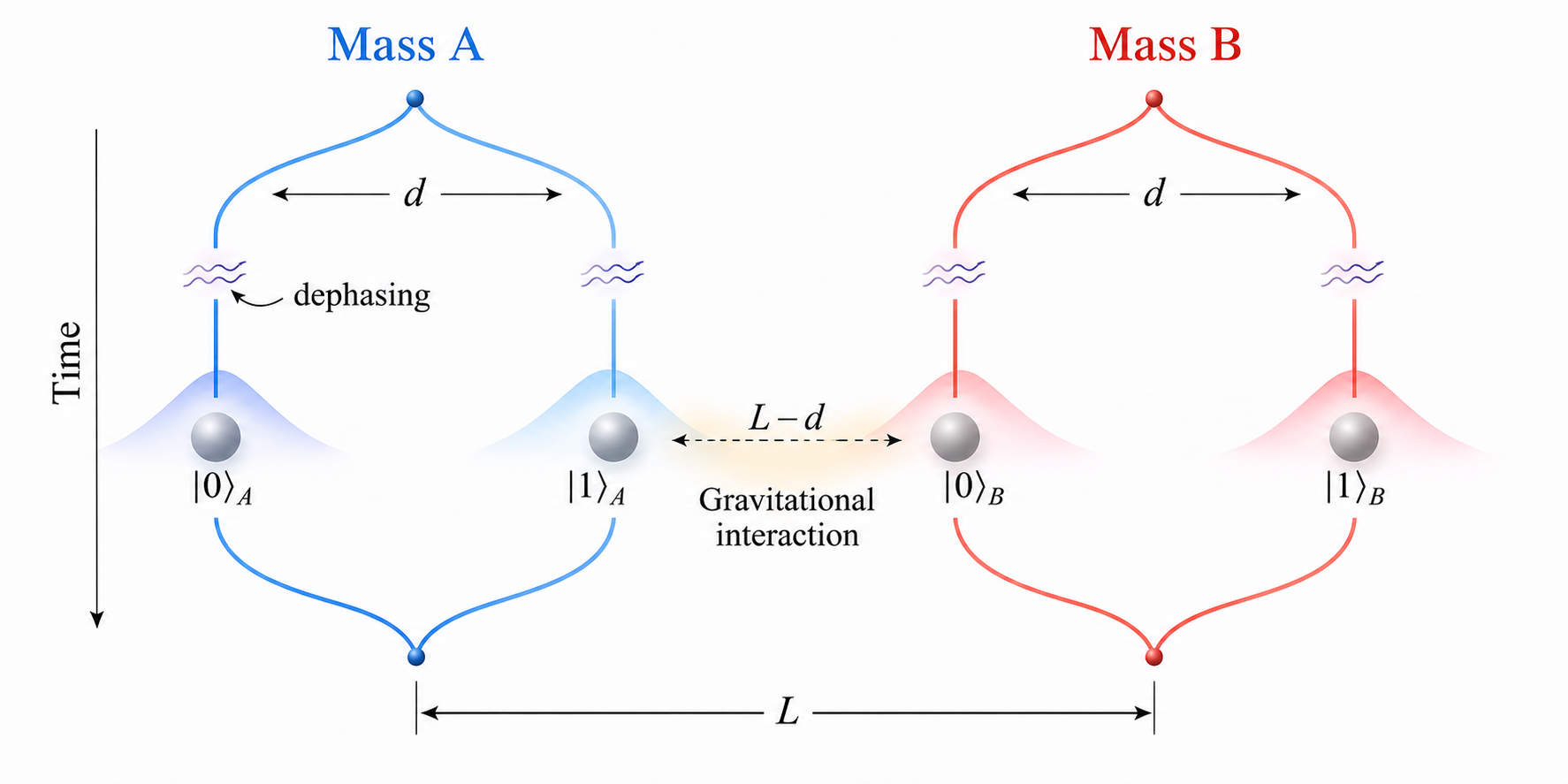}
\caption{\textbf{Gravitational phase accumulation in a dephased BMV interferometer.} Two localized masses $A$ and $B$ are coherently split into path states $|0\rangle$ and $|1\rangle$ separated by $d$, with an interferometer-center separation $L$. In the drawn collinear configuration, the nearest branch distance is $L-d$. The BMV setup enables masses to induce gravitationally mediated entanglement. The purple waves represent local which-path dephasing along the arms, which suppresses path coherence without directly transferring population.}
\label{fig:bmv_setup}
\end{figure*}

This perspective has an important precedent. Lami, Pedernales and Plenio developed local operations and classical communication (LOCC) simulation inequalities for gravitationally induced dynamics and showed that a test of nonclassical mediation need not generate entanglement at any stage \cite{lami2024}. Their prominent example uses ensembles of coherent states of coupled harmonic oscillators and motivates torsion-pendulum implementations \cite{lami2024,carneyStampTaylor2019}. Our setting is distinct. We retain the familiar BMV path-qubit interaction, which is capable of producing entanglement in the absence of noise, and ask whether its conditional dynamical signature persists when local dephasing prevents the usual entanglement witness from succeeding. In other words, the issue is not that the ideal interaction is intrinsically non-entangling. It is that a potentially entangling interaction becomes observationally separable after open-system evolution. This difference permits a direct comparison of a PPT entanglement threshold with independently detectable phase correlations within the same two-path geometry.

For this question, the choice of classical reference dynamics cannot be arbitrary. Ordinary Newtonian interferometry suggests that the gravitational coupling of sufficiently well separated wavepackets is dominated by accumulated action phases and weak forces, not by beam-splitter-like transfer between remote interferometer arms. We quantify that physical no-path-transfer regime with Gaussian overlap and gravitational-deflection estimates, and we identify diagnostics by which unintended arm transfer can be bounded in the experiment. Within this regime, a passive stochastic classical potential acts in each realization by local path phases while the two masses may share classical fluctuations, but a passive source does not use measured path records to condition a phase operation on the remote mass. We call the resulting family path-local (LP) phase channels. Its relevance comes from this physical mechanism and not from mathematical convenience alone.

Negligible arm transfer does not, however, imply that all classical mechanisms are LP channels. A classical controller might measure partial which-path information at one mass and send the outcome to the other, generating active conditional phase responses while preserving populations. Information gain then comes with measurement disturbance. We therefore compare the gravitational dynamics with two nested physical descriptions: passive random phases, for which we obtain an exact finite-time optimum, and a broader population-preserving diagonal PPT class that includes measurement-and-feedback operations. We avoid equating either class with every conceivable classical gravity theory, particularly since interferometric revivals and nonlinear modifications lead to distinct operational questions \cite{streltsov2022,spaventa2023}. Indeed, hybrid, postquantum and collapse-based gravitational models continue to prompt debates about locality, back-action and the operational meaning of classical mediation \cite{diosi1987,penrose1996,tilloyDiosi2016,oppenheim2023,oppenheimDiffusion2023,azizHowl2025,trilloNavascues2025,fengVedralMarletto2026,diosi2025no}.

Our analysis also distinguishes gravitational dynamics from the experimental environment. The same independently characterized which-path dephasing acts in both the quantum and classical descriptions. It is not a freely adjustable resource for a classical simulator. This allows an operational comparison at fixed visibility, with the observed phase-sensitive correlation tied to a concrete dynamical hypothesis. We find an exact trace-distance separation for passive local-phase models even in an all-time separable-output regime, and show that active no-hop measurement and feedback cannot eliminate the leading weak-phase sensitivity within the diagonal PPT class. We then optimize interrogation time, explain the algebraic strong-dephasing scaling, and give a conditional robustness statement for small, separately controlled path-transfer perturbations. The result is neither a device-independent exclusion of all classical gravity nor a replacement for entanglement tests, and it is a physically motivated test of which gravitational phase mechanisms can remain distinguishable after final-state entanglement has ceased to be informative.

\section{Gravitational phase dynamics and entanglement loss}
\label{sec:phase_and_PPT}
The Newtonian ingredient in a BMV interferometer is the same action phase already familiar from single-probe interferometry, but now its source is another delocalized mass. Each pair of paths corresponds to a different separation and therefore a different potential energy. The only phase that cannot be removed using separate local operations on $A$ and $B$ is the alternating combination of the four branch energies. This is the two-body conditional phase that our dynamical test isolates.

Consider two equal masses $A$ and $B$, each with a path qubit $\{|0\rangle,|1\rangle\}$. Let $L$ denote the interferometer center-to-center separation and $d$ the full arm displacement in the collinear geometry. The four branch separations are $R_{00}=R_{11}=L$, $R_{01}=L+d$, and $R_{10}=L-d$, with $L>d$. For a general held geometry, the signed alternating energy combination defines
$\Gamma_g^{\rm(signed)}=\frac{Gm^2}{4\hbar}(R_{00}^{-1}-R_{01}^{-1}-R_{10}^{-1}+R_{11}^{-1})$.
We write $\Gamma_g=|\Gamma_g^{\rm(signed)}|$ and choose the logical labeling of one path qubit so that the interaction has the sign convention stated below. In the collinear geometry shown here, the signed alternating combination is negative before this relabeling. For time-dependent trajectories, one integrates the corresponding signed energy combination along the protocol. None of our channel-level statements depends on the collinear geometry except through the resulting phase $\Theta$. After subtracting an overall phase and phases local to the two interferometers, the collinear interaction takes the form
\begin{equation}
\begin{aligned}
 H_g&=-\hbar\Gamma_g Z_A Z_B,\\
 \Gamma_g&=\frac{Gm^2}{4\hbar}
 \left(\frac1{L-d}+\frac1{L+d}-\frac2L\right).
\end{aligned}
\label{eq:grav_rate}
\end{equation}
We choose path labels so that $\Gamma_g\ge0$ and write $\Theta=\Gamma_g\tau$ and $U_\Theta=e^{i\Theta Z_AZ_B}$. A held-separation geometry is assumed in Eq.~\eqref{eq:grav_rate}. A realistic pulse sequence replaces the constant separations by trajectory-dependent distances integrated over time.

In an ideal phase experiment, branch populations are approximately unchanged during the hold, but coherences decay when the environment acquires information about the occupied path. The effective qubit model describes this ordinary which-path coupling with a local dephasing rate. It does not identify the microscopic origin of every noise process, and in particular it does not treat the environment as a free resource available to a proposed gravitational mediator. We take the same independently characterized, population-preserving Markovian dephasing on both masses:
\begin{equation}
\dot\rho=-\frac{i}{\hbar}[H_g,\rho]
+\gamma\sum_{j=A,B}(Z_j\rho Z_j-\rho).
\label{eq:master}
\end{equation}
The Hamiltonian and dissipators commute as superoperators. Therefore the exact quantum map is $\mathcal E_Q(\tau)=\Dd_r\circ\cU_\Theta$, where $\cU_\Theta(\rho)=U_\Theta\rho U_\Theta^\dagger$, the single-particle coherence visibility is $r=e^{-2\gamma\tau}$, and $\Dd_r=\Dd_r^A\otimes\Dd_r^B$.
For the product input $\rho_1=|++\rangle\langle++|$, with $|+\rangle=(|0\rangle+|1\rangle)/\sqrt2$, the output reads
\begin{equation}
\sigma_1=\frac14
\begin{pmatrix}
1&r e^{i2\Theta}&r e^{i2\Theta}&r^2\\
r e^{-i2\Theta}&1&r^2&r e^{-i2\Theta}\\
r e^{-i2\Theta}&r^2&1&r e^{-i2\Theta}\\
r^2&r e^{i2\Theta}&r e^{i2\Theta}&1
\end{pmatrix}.
\label{eq:sigma_plus}
\end{equation}
The conditional phase tends to generate entanglement while the dephasing reduces the off-diagonal elements on which it depends. Their competition can be characterized exactly, without inferring entanglement from visibility alone. For a two-qubit output, positivity of the partial transpose is equivalent to separability \cite{peres1996,horodeckiPPT1996}. The potentially negative partial-transpose eigenvalue is
\begin{equation}
\lambda_{\rm ent}=\frac14\left[1-r^2-2r|\sin2\Theta|\right].
\label{eq:ppt_ent_eigenvalue}
\end{equation}
Since the state is two-qubit, the PPT criterion is necessary and sufficient for separability. Consequently, $\sigma_1$ is separable exactly when
\begin{equation}
1-r^2\ge2r|\sin2\Theta|,
\label{eq:separability_threshold}
\end{equation}
or $\gamma\ge\frac{\operatorname{arsinh}(|\sin2\Theta|)}{2\tau}$.
For $\Theta=\Gamma_g\tau$ we have $\operatorname{arsinh}(|\sin2\Theta|)\le2\Gamma_g\tau$, so $\gamma\ge\Gamma_g$ guarantees that the output from $|++\rangle$ is separable for all $\tau>0$. This assertion concerns the stated input, not every output of the two-qubit channel. One can find details in Appendix~\ref{app:exact_dynamics} and Appendix~\ref{app:eigenvalues}.

\section{Physical basis of the no-path-transfer approximation}
\label{sec:nohop}
There are two distinct notions that are sometimes conflated in discussions of gravitational interferometry. The first is phase accumulation: a wavepacket following a prescribed arm acquires an action-dependent phase, just as in COW neutron interferometry and more recent gravitational Aharonov--Bohm-type atom interferometry \cite{cow1975,overstreet2022}. The second is motion: the same potential exerts a force, changes the packet centre and can modify the interferometer contrast. Neither is, by itself, a coherent beam-splitter-like transition from one macroscopically separated arm to the other. The fact that gravitational phases can be extracted while path deflections are characterized separately provides direct experimental motivation for keeping these effects conceptually distinct.

A conservative Newtonian potential acts by multiplication in the position representation, $V(\hat x)\psi(x)=V(x)\psi(x)$. For prescribed, well-resolved paths, the action phase is therefore the leading contribution to the projected path evolution. An off-diagonal matrix element between remote wavepackets is additionally penalized by their spatial overlap. This is the physical basis for an approximately population-preserving, or ``no-hop'', path description. It remains an approximation: free evolution, wavepacket deformation and gravitational acceleration must be bounded independently of the projected off-diagonal matrix element. We quantify these scales rather than assume them negligible by definition.

To estimate how strongly a smooth potential could couple the two resolved path modes, consider localized Gaussian packets whose centre separation is much greater than their width. The spatial overlap is exponentially suppressed even when the Newtonian potential itself is not exceedingly small. This makes the off-diagonal transition effect parametrically different from the branch action phase, which remains proportional to the potential-energy difference and the interrogation time. For two normalized one-dimensional Gaussians of rms width $\sigma_x$ and separation $d$, the overlap is $S=\exp[-d^2/(8\sigma_x^2)]$. A bounded smooth residual potential $V$ has an off-diagonal overlap matrix element controlled by $V_*S$. Because nonorthogonal Gaussians do not define an exact qubit basis, the correct two-mode estimate uses a Gram--Schmidt orthonormalization. If $|V(x)|\le V_*$ on the relevant support, the orthonormalized matrix element obeys
\begin{equation}
|\langle e_1|V|e_0\rangle|
\le \frac{2V_*S}{\sqrt{1-S^2}}.
\label{eq:orthogonal_hop_bound}
\end{equation}
A first-order two-mode perturbative estimate consequently gives
\begin{equation}
P_{\rm hop}^{\rm(est)}\lesssim
\left(\frac{2V_*\tau}{\hbar}\right)^2
\frac{e^{-d^2/(4\sigma_x^2)}}{1-e^{-d^2/(4\sigma_x^2)}}.
\label{eq:hop_prob_bound}
\end{equation}
This is an estimate within a prescribed two-mode model, not a universal operator-norm bound on unrestricted free wavepacket dynamics. The complete assumptions and orthonormalization are set out in Appendix~\ref{app:path_transfer}.

\begin{figure*}[t]
\centering
\includegraphics[width=0.93\textwidth]{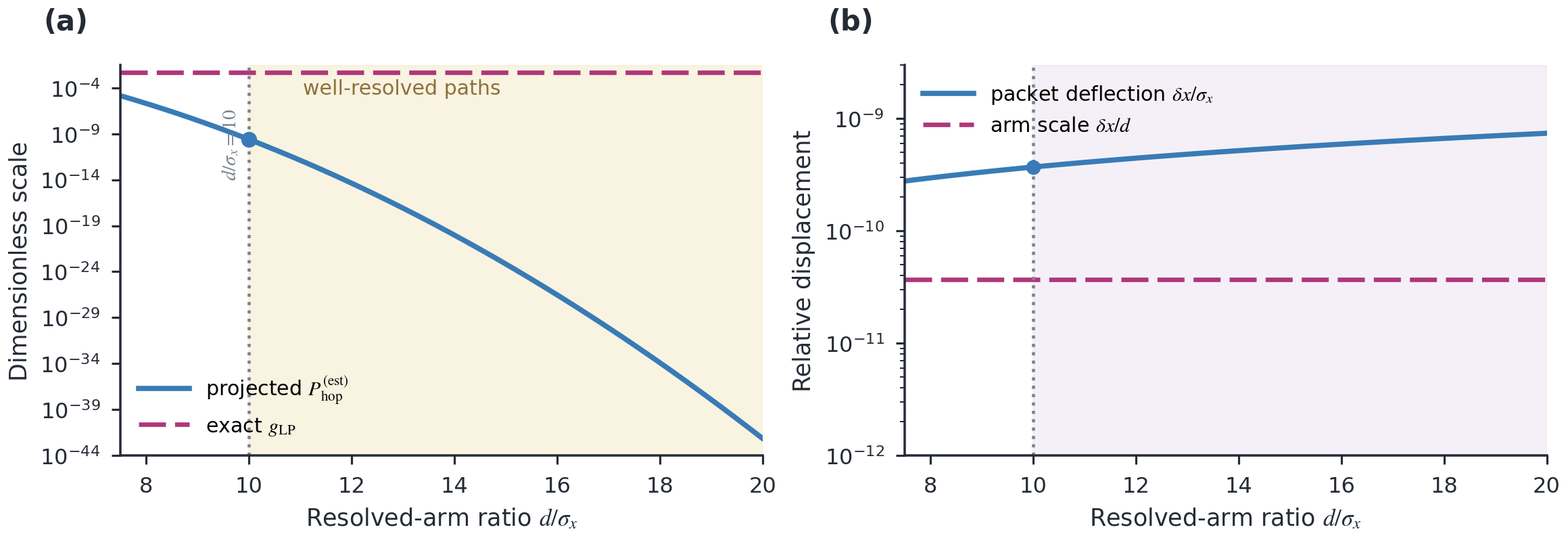}
\caption{\textbf{Quantitative scales supporting a no-path-transfer approximation.} The illustrative parameters are $m=10^{-14}\,\mathrm{kg}$, $L=40\,\mu\mathrm m$, $d=10\,\mu\mathrm m$ and $\tau=1\,\mathrm s$, chosen within the order-of-magnitude ranges discussed in the original BMV proposal and recent experimental roadmaps \cite{bose_spin_2017,boseRoadmap2025,boseWhitepaper2026}. The $30\,\mu\mathrm m$ minimum branch separation is an illustrative benchmark for an electromagnetically shielded design, rather than an unshielded gravity-dominated configuration. Residual particle–shield interactions and shield-induced force noise require independent control. The assumed visibility $r=0.1$ and varied Gaussian width $\sigma_x$ are model choices, not experimental measurements. (a) The perturbative two-mode transfer estimate in Eq.~\eqref{eq:hop_prob_bound} and exact passive LP gap (see \autoref{thm:LP_phase_gap}) versus $d/\sigma_x$, and the shaded region denotes well-resolved packets. The conservative energy scale is $V_*=Gm^2/(L-d)$. (b) The associated Newtonian displacement relative to packet width and arm separation. These quantities are distinct and do not by themselves yield a full-channel error bound or establish experimental feasibility.}
\label{fig:nohop_quantitative}
\end{figure*}

The overlap estimate controls a projected matrix element, and it is not a complete analysis of freely moving wavepackets. In particular, gravity can change the trajectory smoothly without causing any discrete path flip. For this separate effect, a conservative bound on Newtonian trajectory deflection follows from the minimum intermass distance $R_{\min}=L-d$:
\begin{equation}
\delta x\lesssim\frac{Gm\tau^2}{2R_{\min}^2}.
\label{eq:gravitational_displacement}
\end{equation}
The ratios $\delta x/d$ and $\delta x/\sigma_x$ respectively probe inter-arm transfer and deviations from a prescribed wavepacket trajectory. Both effects are independently accessible through single-arm characterization and path-population diagnostics. For the illustration we select a BMV-scale geometry informed by the original proposal and recent community roadmaps, which consider masses near $10^{-14}\,\mathrm{kg}$, tens of micrometres in separation, micrometre-scale path splitting and interaction times up to approximately one second \cite{bose_spin_2017,boseRoadmap2025,boseWhitepaper2026}. These ranges motivate the mass, geometry and hold time, not the Gaussian packet width or the independently assumed visibility. Figure~\ref{fig:nohop_quantitative} illustrates the corresponding estimates and their dependence on path resolution.

The two numerical estimates can be checked experimentally using arm-resolved population readout, single-interferometer visibility measurements and tests of source-induced deflection. Their role is to identify a parameter region where path transfer and trajectory shifts are much smaller than the phase signature, and they do not substitute for calibration of electromagnetic coupling, loss or non-phase noise. This is precisely why one should not infer a model-independent gravity conclusion from the overlap integral alone.

The phase-and-force picture is supported by gravity-sensitive neutron and atom interferometry, in which measurable phase shifts do not require gravitationally driven transfer between interferometer arms \cite{cow1975,abeleleeb2012,overstreet2022}. The micromechanical force measurements and molecular-interference experiments cited above also emphasize why a small gravitational acceleration and a preserved coherent path basis are different experimental constraints \cite{westphal2021,fein2019}. The no-hop approximation is therefore not introduced merely to simplify the channel problem: it describes the regime where Newtonian gravity acts predominantly through phases and small forces, rather than as a beam splitter between macroscopically separated paths. The scope of the classical model must nevertheless be specified carefully. A passive classical random-potential description with no access to path measurement records produces, in each realization, local phase factors $U_A(\lambda)\otimes U_B(\lambda)$, and shared randomness can correlate the two local phases. This is the path-local (LP) class defined below. A general no-hop transformation need not belong to LP: the target quantum conditional-phase gate $U_\Theta$ is itself diagonal, and active classical measurement and feedback can produce conditional coherence effects without changing populations. Those operations are treated by the broader diagonal-PPT comparison in Sec.~\ref{sec:active_nohop_feedback}. Neither overlap suppression nor the Choi benchmark alone is used to claim that all classical gravity theories reduce to LP.

\section{Exact discrimination beyond final-state entanglement}
\label{sec:phase_dynamics_test}
What does a passive classical description predict within this phase-dominated regime? If a classical fluctuating potential supplies only a random background for the two interferometers, then each run assigns a phase to each arm of each mass. The same fluctuating source may correlate these phase shifts across the laboratories, but without measuring a path variable and feeding its result to the other laboratory it does not implement a coherent phase conditional on both path qubits. The operational class capturing this passive mechanism is a convex mixture of local diagonal unitary phases. We allow the shared randomness to be completely general rather than impose independent phase noise, so the comparison does not artificially exclude classical correlations.

The passive path-local phase class consists of all channels
\begin{equation}
\mathcal C_{\rm LP}(\rho)=\int d\lambda\,p(\lambda)
[U_A(\lambda)\otimes U_B(\lambda)]\rho
[U_A^\dagger(\lambda)\otimes U_B^\dagger(\lambda)],
\label{eq:LP_definition_clean}
\end{equation}
where $U_j(\lambda)=e^{i\alpha_{j,0}(\lambda)}|0\rangle\langle0|+e^{i\alpha_{j,1}(\lambda)}|1\rangle\langle1|$, and the distribution $p(\lambda)$ is allowed to correlate the two interferometers. A further physical restriction is that the laboratory background dephasing is not reoptimized as part of whichever mediator is chosen. Both hypotheses describe the same apparatus and hence share the same measured visibility loss. This separation is essential: allowing a simulator to prepare only the already-dephased final state would erase the distinction between the gravitational interaction and its environment. With the same external dephasing applied to both descriptions, the resulting finite-time comparison is
\begin{equation}
\begin{split}
\sigma_1&=\Dd_r\circ\cU_\Theta(\rho_1),\\
\tilde\sigma_1&=\Dd_r\circ\mathcal C_{\rm LP}(\rho_1).
\end{split}
\label{eq:channel_benchmark}
\end{equation}
This also describes interleaved passive LP steps and local dephasing, because each LP step commutes with $\Dd_r$ and compositions of LP channels remain LP, i.e., $
(\Dd_{r_N}\circ\mathcal C_N)\circ\cdots\circ
(\Dd_{r_1}\circ\mathcal C_1)=
\Dd_{\prod_j r_j}\circ (\mathcal C_N\circ\cdots\circ\mathcal C_1)
$. The finite-time gap is the minimum trace distance
\begin{equation}
g_{\rm LP}(r,\Theta)=\inf_{\mathcal C_{\rm LP}}
\frac12\|\sigma_1-\Dd_r\circ\mathcal C_{\rm LP}(\rho_1)\|_1.
\label{eq:LP_gap_definition}
\end{equation}

The measurable distinction can be understood without introducing a process matrix. Condition on whether $A$ occupies path $0$ or path $1$, and examine a coherence of the other mass $B$. The Newtonian $Z_AZ_B$ interaction rotates these two conditional coherences in opposite directions, because the effective gravitational potential experienced by $B$ depends on $A$'s branch. Ordinary local dephasing attenuates both coherences by the same factor, and it does not remove their opposing phase orientation. Explicitly, the gravitational output contains
\begin{equation}
(\sigma_1)_{00,01}=\frac r4e^{+i2\Theta},\qquad
(\sigma_1)_{10,11}=\frac r4e^{-i2\Theta}.
\label{eq:grav_conditional_coherences}
\end{equation}
A passive local random-phase model has no access to the conditional branch information required to reverse the phase between these two sectors. In particular, averaging over shared random local phases may attenuate the coherence, but it produces the same $B$-coherence factor in the two $A$ sectors. This is a physically transparent equality rather than a consequence of optimizing a semidefinite program. An interferometric recombination measuring a local $Y_B$ quadrature and correlating it with the path label $Z_A$ directly addresses the discrepancy. The corresponding normalized Hermitian observable is
\begin{equation}
W_B=-Z_A\otimes Y_B,
\label{eq:witness_WB_main}
\end{equation}
with $\|W_B\|_\infty=1$
obeys $\langle W_B\rangle_{\sigma_1}=r\sin2\Theta$ and $\langle W_B\rangle_{\tilde\sigma_1}=0$. Trace-distance duality gives $g_{\rm LP}\ge\frac r2|\sin2\Theta|$. We now strengthen this witness inequality to an exact optimum. The trace-distance lower bound supplied by this witness
equals the minimum distance over the entire passive
random-phase class.

\begin{theorem}[Exact finite-time LP simulation gap]
\label{thm:LP_phase_gap}
For the coherent product probe $\rho_1=|++\rangle\langle++|$, arbitrary $0\le r\le1$, and $U_\Theta=e^{i\Theta Z_AZ_B}$, the optimum with the same external dephasing applied to both descriptions, over all passive path-local phase channels, is
\begin{equation}
g_{\rm LP}(r,\Theta)=\frac r2|\sin2\Theta|.
\label{eq:main_LP_gap}
\end{equation}
An optimal channel at each specified interrogation time is
$\mathcal C_*(\rho)=\cos^2\Theta\,\rho+
\sin^2\Theta\,(Z_AZ_B)\rho(Z_AZ_B).$
\end{theorem}

\noindent Proof. The witness $W_B$ proves the lower bound. For $\mathcal C_*(\rho)$, direct substitution into Eq.~\eqref{eq:channel_benchmark} gives
\begin{equation}
\Delta_*:=\sigma_1-\Dd_r\circ\mathcal C_*(\rho_1)
=-\frac{r\sin2\Theta}{4}(Y_AZ_B+Z_AY_B).
\label{eq:LP_optimal_difference}
\end{equation}
Since $Y_AZ_B$ and $Z_AY_B$ commute, square to the identity and have product $X_AX_B$, their sum has spectrum $\{2,-2,0,0\}$. Thus the nonzero eigenvalues of $\Delta_*$ are $\pm r\sin2\Theta/2$. Hence $\|\Delta_*\|_1/2=r|\sin2\Theta|/2$, saturating the witness bound. \(\square\)

The optimal passive channel has an intuitive interpretation: with suitable shared classical probabilities, both interferometers either receive the same local $Z$ flip or neither receives one. Such correlated phase noise cannot reproduce the conditional phase itself, but it is the closest possible passive approximation for this probe and metric. The optimum holds at each chosen interrogation time, not necessarily for one generator that is simultaneously optimal at all times. For $0\le|\Theta|<\pi/4$, a Poisson process of correlated local $Z$ flips realizes the optimal terminal channel with rate $\kappa=-\ln(\cos2\Theta)/(2\tau)$, and the associated generator is
$
\mathcal L_*(\rho)=\kappa[(Z_AZ_B)\rho(Z_AZ_B)-\rho],
$
Here the rate depends on the selected $\tau$, and no fixed stochastic law is claimed to be jointly optimal for every interrogation time. The construction and boundary behavior are detailed in Appendix~\ref{app:LP_proof}.

A useful independent check is furnished by the normalized Choi trace distance. In the absence of dephasing, the same random phase mixture achieves the full PPT-Choi lower bound:
\begin{equation}
\begin{split}
\inf_{\Lambda\in{\rm LP}}D_{\rm Choi}(\cU_\Theta,\Lambda)
&=\inf_{\Lambda\in{\rm PPT\text{-}Choi}}D_{\rm Choi}(\cU_\Theta,\Lambda)\\
&=\frac{|\sin2\Theta|}{2}.
\end{split}
\label{eq:choi_equivalence_main}
\end{equation}
One can find the proof in Appendix~\ref{app:choi_equivalence}. 

\begin{figure*}[t]
\centering
\includegraphics[width=0.96\textwidth]{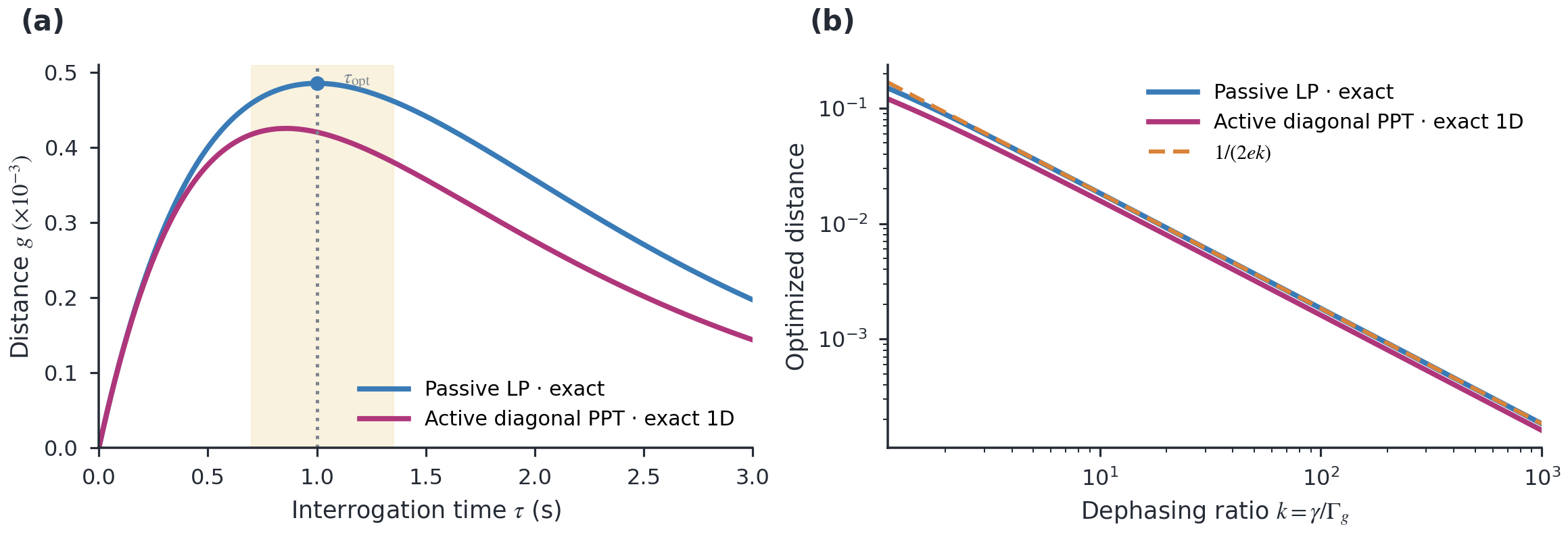}
\caption{\textbf{Exact passive and diagonal-PPT simulation gaps.} (a) Gaps at $\Gamma_g=1.32\times10^{-3}\,\mathrm{s}^{-1}$ and $\gamma=0.5\,\mathrm{s}^{-1}$, for which $\gamma>\Gamma_g$ and the output state of $|++\rangle$ remains PPT at every time. The passive curve evaluates the analytic exact distance, while the active curve numerically minimizes the exact one-dimensional diagonal-PPT objective given in Appendix~\ref{app:diag_ppt_proof}, and neither curve substitutes the small-phase approximation or global analytic lower bound. The dotted line indicates $\tau_{\mathrm{opt}}$ for LP. (b) Optimized distances versus $k=\gamma/\Gamma_g$, compared with the passive strong-noise asymptote. These model-dependent distinguishability measures
provide a quantitative basis for experimental tests
under the stated assumptions.}
\label{fig:strong_dephasing_gaps}
\end{figure*}

\section{Beyond passive classical mediation}
\label{sec:active_nohop_feedback}
The phase-only description is physically natural for a passive background potential, but it does not exhaust classical control. A controller may acquire partial information about the path of mass $A$, send a classical record to mass $B$ and adjust its phase. Such a protocol can generate branch-conditioned phase responses without directly transferring either mass between arms. The missing ingredient is that obtaining path information disturbs the conjugate coherence: stronger information gain generally entails greater back-action. The question is whether this trade-off permits active classical feedback to erase the discrepancy left by passive random potentials.

The physics of measurement-induced back-action is already central to classical-channel gravity models: a gravitational force implemented through continuously measured positions and classical records is accompanied by irreducible noise \cite{Kafri_2014,kafri2015,tilloyDiosi2016}. We do not equate all such models, but this literature makes it clear why an analysis restricted to uncontrolled random phases would leave an important alternative untested.

We address that question by comparing with a larger, mathematically tractable family of population-preserving diagonal channels. 
For two path qubits, the Schur correlation matrix is a
$2\times2$ bipartite operator, so positivity under partial
transpose is equivalent to separability. The feasible set
therefore coincides with the diagonal separable-channel
class and contains every diagonal LOCC channel.
Explicitly, the Schur channel map is given as
\begin{equation}
[\mathcal F_M(\rho)]_{xy}=M_{xy}\rho_{xy},
\label{eq:diagPPT_schur_def}
\end{equation}
with $\quad M\succeq0,$ $ M_{xx}=1$, and $\quad M^{\Gamma_B}\succeq0.$
Here $x,y\in\{00,01,10,11\}$ and $\Gamma_B$ denotes partial transpose with respect to the $B$ index in the $2\times2$ path-label bipartition of the Schur matrix. These constraints provide an exact characterization of
diagonal separable channels. The constraint is on the Schur correlation matrix $M$, not on the final two-mass state. Define
\begin{equation}
\begin{split}
g_{\rm diagPPT}(r,\Theta)&=\inf_{M\in{\rm diagPPT}}\\
&\quad D\!\left(\Dd_r\circ\cU_\Theta(\rho_1),
\Dd_r\circ\mathcal F_M(\rho_1)\right).
\end{split}
\label{eq:g_diagPPT_def}
\end{equation}
We have the following theorem for no-hop models.

\begin{theorem}[Diagonal no-hop feedback gap]
\label{thm:diagPPT_gap}
For $0<r\le1$, write $s_r=\sqrt{2-r^2}$. The diagonal-PPT gap obeys
\begin{equation}
\begin{split}
g_{\rm diagPPT}(r,\Theta)\ge\frac r4\big[&(s_r+r)|\sin2\Theta|\\
&-(s_r-r)(1-\cos2\Theta)\big].
\end{split}
\label{eq:diagPPT_global_bound}
\end{equation}
In the weak-phase regime, the tight leading behavior is
$
g_{\rm diagPPT}(r,\Theta)=
\frac r2(r+\sqrt{2-r^2})|\Theta|+O(\Theta^2).
$
\end{theorem}
Moreover, this optimum is attained by a diagonal LOCC
channel, so
$g_{\rm diagLOCC}=g_{\rm diagSEP}=g_{\rm diagPPT}$. 
The physical point of this result is the persistence, not the numerical universality, of phase sensitivity. Passive random phases cannot create the opposite conditional coherences at all, and active classical feedback can partly imitate them, but must pay an information--disturbance cost. This cost reduces the optimal coefficient in the small-phase regime without changing its linear order. The full proof, exact one-dimensional variational
representation, and a finite-round diagonal LOCC protocol
attaining the optimum are given in Appendix~\ref{app:diag_ppt_proof}. The right-hand side of Eq.~\eqref{eq:diagPPT_global_bound} can be negative outside the useful weak-phase range, and in that case the trivial bound $g_{\rm diagPPT}\ge0$ should be used. The ratio of the active coefficient to the passive coefficient is $(r+\sqrt{2-r^2})/2$, between $1/\sqrt2$ and $1$. Thus the broader diagonal class reduces the leading constant but cannot eliminate its linear phase dependence. This does not by itself prove an identical bound against unrestricted non-diagonal LOCC or arbitrary time-adaptive measurement-feedback combs.

\section{Optimal interrogation time and strong-dephasing scaling}
\label{sec:scaling}
A weak gravitational interaction favors long interrogation times because the conditional phase grows linearly with the duration of the hold. Environmental dephasing favors the opposite strategy: late-time phase information is exponentially attenuated. Our exact gap permits this trade-off to be optimized without replacing the classical comparison by a heuristic witness lower bound. For fixed $\Gamma_g$ and $\gamma$, the exact passive distance in Theorem~\ref{thm:LP_phase_gap} gives
\begin{equation}
g_{\rm LP}(\tau)=\frac12e^{-2\gamma\tau}|\sin(2\Gamma_g\tau)|.
\label{eq:LP_gap_vs_time}
\end{equation}
Writing $k=\gamma/\Gamma_g$ and $y=2\Gamma_g\tau$, the first positive lobe maximizes $e^{-ky}\sin y/2$, and all later lobes are smaller for $\gamma>0$. The exact optimum for $\gamma>0$ is
\begin{align}
\tau_{\rm opt}&=\frac1{2\Gamma_g}\arctan\frac{\Gamma_g}{\gamma},
\label{eq:exact_tau_opt}\\
g_{\rm LP}^{\max}&=\frac{\Gamma_g}{2\sqrt{\gamma^2+\Gamma_g^2}}\exp\!\left[-\frac{\gamma}{\Gamma_g}\arctan\frac{\Gamma_g}{\gamma}\right].
\label{eq:exact_gap_opt}
\end{align}
At $\gamma=0$ the first maximum is at $\tau=\pi/(4\Gamma_g)$ and the value is $1/2$. For $\gamma\gg\Gamma_g$,
$
\tau_{\rm opt}\sim\frac1{2\gamma}$ and $
g_{\rm LP}^{\max}\sim\frac1{2e}\frac{\Gamma_g}{\gamma}.
$
The optimum occurs when the experiment has accumulated enough conditional phase to be visible, but before the environment has erased most of the local coherence. In particular, $\tau_{\rm opt}$ approaches the dephasing timescale $1/(2\gamma)$ in the strongly noisy regime. Hence the optimized passive-model separation has an algebraic noise penalty rather than requiring a long interaction after coherence has vanished. This is a property of the finite-time comparison with external dephasing held fixed, not of a particular estimator's sample complexity or a guarantee of experimentally achievable sensitivity.

For the active diagonal-PPT class, for small phase, $
g_{\rm diagPPT}(r,\Theta)=
\frac r2(r+\sqrt{2-r^2})|\Theta|+O(\Theta^2)
$ gives a different visibility-dependent coefficient. Its exact objective can be optimized over time via the one-dimensional reduction in Appendix~\ref{app:diag_ppt_proof}. Figure~\ref{fig:strong_dephasing_gaps} uses this exact reduced minimization, rather than replacing the active result with the weaker analytic lower-bound curve. Both curves show the same algebraic $\Gamma_g/\gamma$ strong-noise trend for the tested model classes.

\section{Low-visibility BMV estimate and calibration requirements}
\label{sec:experiment}
To give the distances an operational scale, consider a fixed geometry and a separately characterized visibility. The gravitational phase can be estimated from the known masses, trajectories and interaction duration. The relevant phase quadrature can be extracted by recombining the interferometer arms and measuring local Pauli correlations. These steps assume that competing interactions capable of producing correlated phases have been bounded. For an illustrative held geometry $m=10^{-14}\,\mathrm{kg}$, $L=40\,\mu\mathrm m$, $d=10\,\mu\mathrm m$, and $\tau=1\,\mathrm s$, Eq.~\eqref{eq:grav_rate} gives $\Theta=0.052741\,\mathrm{rad}$ \cite{bose_spin_2017,boseRoadmap2025,boseWhitepaper2026}. Taking the independently measured single-qubit visibility to be $r=0.1$, the observable $W_B=-Z_A Y_B$ has expectation
\begin{equation}
S_B:=\langle W_B\rangle_{\sigma_1}=r\sin2\Theta=1.05286\times10^{-2}.
\label{eq:low_visibility_signal}
\end{equation}
The final state is PPT and separable, whereas the exact passive LP distance is
\begin{equation}
 g_{\rm LP}=5.2643\times10^{-3}.
\label{eq:low_visibility_gap}
\end{equation}
For the same parameters, the diagonal-PPT global inequality gives a separate, weaker certificate, and its right-hand side should not be confused with the exact diagonal-PPT optimum. The numerical examples in Figs.~\ref{fig:nohop_quantitative} and \ref{fig:strong_dephasing_gaps} use different parameters to isolate distinct issues: physical wavepacket scales and asymptotic strong-dephasing behavior, respectively.

This is a deliberately simplified phase-qubit estimate rather than an experimentally established gravity-mediated signal. Existing efforts in matter-wave splitting, spin-dependent forces and mesoscopic control face practical visibility and background-interaction constraints well beyond the two-level dephasing model \cite{boseRoadmap2025,carneyStampTaylor2019,pedernales2020,vandeKamp2020}. The quantitative comparison makes a different point: the dephasing factor need not vanish for a distinguishable conditional phase to survive mathematically.

The measurement of a nonzero $S_B$ is a test of the passive LP model only when local phases, detection imbalance, residual non-gravitational $Z_AZ_B$ interactions, and the environmental visibility have been independently characterized. A direct measurement of $W_B$ alone does not certify the active diagonal-PPT bound because active no-hop feedback may generate a nonzero $W_B$ expectation. Certifying that broader gap requires the full trace-distance discrimination measurement, or the appropriate multi-observable dual witness from Appendix~\ref{app:diag_ppt_proof}. A single $\pm1$ Pauli-product measurement has standard error at most $1/\sqrt N$ for $N$ independent samples, before accounting for calibration biases. With a desired statistical significance $z$, the idealized scaling is $N\gtrsim z^2/S_B^2$. Here $z=5$ requires roughly $2.3\times10^5$ independent detected events before losses, confidence-procedure corrections and nuisance-parameter uncertainties are included.

The event-count estimate is deliberately optimistic: correlated drift, unequal detector efficiencies, arm-dependent attenuation and uncertainty in the measured visibility can require substantially more data. Conversely, the target is not full two-mass state tomography if the passive model is the declared hypothesis class, and the simple conditional-coherence witness suffices for that particular rejection test. The broader active feedback class requires multiple observables or a dual-optimal measurement, rather than relying on a single phase-quadrature correlation.

A diagnostic input $\rho_2=(|00\rangle\langle00|+|11\rangle\langle11|)/2$ is invariant under the target phase gate and all no-hop Schur channels. Leakage into $\Pi_{\rm odd}=|01\rangle\langle01|+|10\rangle\langle10|$ under a more general simulator is directly quantified by $p_{\rm leak}=\mathrm{Tr}(\Pi_{\rm odd}\tilde\sigma_2)$, implying $D(\rho_2,\tilde\sigma_2)\ge p_{\rm leak}$. This is a population diagnostic, not a proof that every path-hopping simulator is excluded by these two inputs. Extensions to asymmetric, non-Markovian phase noise and bounded channel-calibration errors appear in Appendix~\ref{app:general_noise}.

\section{Robustness to finite path transfer}
\label{sec:robustness_hopping}
The no-hop approximation is strongest when it is testable rather than assumed to be exact. In a real interferometer, a weak residual off-diagonal coupling or imperfect control can introduce a small amplitude for transitions between the nominal arms. Such imperfections should not be dismissed simply because the ideal Newtonian phase dominates. Equally, a measured leakage probability cannot be subtracted directly from a trace-distance gap: transition probabilities are quadratic in small amplitudes, whereas the worst-case change of a state can be linear in those amplitudes.

A useful stability statement follows from the triangle inequality. Let $\sigma_Q$ denote the ideal dephased quantum output and $\sigma_Q^{\rm phys}$ its experimentally realized counterpart. Assume $D(\sigma_Q^{\rm phys},\sigma_Q)\le\epsilon_Q$. For every simulator in a physically specified family $\mathfrak C_{\rm phys}$, suppose there exists a passive LP simulator whose dephased output differs by at most $\epsilon_C$, uniformly over the family. The exact LP theorem then implies
\begin{equation}
\begin{split}
g_{\rm phys}&:=\inf_{\mathcal C\in\mathfrak C_{\rm phys}}
D\bigl(\sigma_Q^{\rm phys},
\mathcal D_r\circ\mathcal C(\rho_1)\bigr)\\
&\ge \max\left\{0,\frac r2|\sin2\Theta|
-\epsilon_Q-\epsilon_C\right\}.
\end{split}
\label{eq:finite_hop_robustness_main}
\end{equation}
This result quantifies how much model error the witness can tolerate, but its force depends on establishing the two error bounds independently. It does not extend the passive theorem to all no-hop feedback models or to unrestricted classical gravitational dynamics.

There is a concrete sufficient condition for $\epsilon_C$ in a microscopic passive model. Suppose the model has a reference evolution consisting of local diagonal phases and the common dephasing, and that its only extra interaction is a residual Hermitian path-changing Hamiltonian $H_{\rm off}(t)$. The Duhamel expansion and trace-norm contractivity of the reference completely positive evolution give
\begin{equation}
\epsilon_C\le\min\!\left\{1,\frac1\hbar\int_0^\tau\!dt\,\|H_{\rm off}(t)\|_\infty\right\},
\label{eq:hop_amplitude_bound_main}
\end{equation}
The same bound must hold across the shared-randomness realizations (uniformly or in expectation). This is an amplitude-level dynamical bound, not an equality between the probability of finding a particle in the wrong arm and the error of the entire channel. The Gaussian and acceleration estimates in Fig.~\ref{fig:nohop_quantitative} motivate why the integral can be small under controlled wavepacket assumptions, while arm-population tests provide a separate experimental consistency check. The derivation and its limitations are given in Appendix~\ref{app:path_transfer}.

\section{Discussion and outlook}
\label{sec:discussion}
The historical success of gravity-sensitive interferometry gives a useful perspective on our result. COW-type experiments show that gravity can imprint a measurable phase on a spatially coherent quantum probe in a regime where momentum transfer between macroscopic paths is not the operative signal. The BMV setting asks what changes when that phase is sourced by a second coherent quantum mass \cite{bose_spin_2017,marletto_gravitationally_2017}. A conditional phase is invisible to an entirely passive collection of local random action phases, even after ordinary which-path decoherence makes the two-probe output separable. This is the physical origin of the channel discrimination observable.

Physical modelling is essential to the interpretation. The localized-wavepacket estimates support negligible direct arm transfer and tiny gravitational deflection in a parameter regime relevant to phase interferometry, but these estimates cannot turn the no-hop assumption into a theorem about all classical gravity. More specifically, no-hop dynamics permits conditional coherence manipulation, and passive random potentials form a strict subset of the no-hop channel class. The diagonal-PPT calculation establishes a nonzero gap even after admitting a broad class of active population-preserving separable operations, and this is the model-robustness contribution of our second theorem. What it does not establish is the same positive noisy gap against arbitrary adaptive LOCC strategies without additional dynamical hypotheses.

A key experimental requirement is that local which-path dephasing be characterized independently and applied identically to the quantum and classical hypotheses. Classical measurement-and-feedback models often generate their own irreducible back-action, which is part of the proposed mediator model and cannot simply be interchanged with a separately calibrated background \cite{kafri2015,pfister2016,tilloyDiosi2016}. Our diagonal no-hop PPT comparison permits local path measurements and classically communicated feedback at the reduced path-qubit level. 
The associated measurement back-action is part of the optimized classical comparison. 
Kafri–Taylor–Milburn-type measurement-and-feedback models are genuine classical-communication implementations, with unavoidable measurement-induced noise \cite{kafri2013noiseinequalityclassicalforces,Kafri_2014}. 
When such measurement-and-feedback dynamics is reduced to a
population-preserving diagonal path-qubit channel, its Schur
description is included in our active diagonal comparison,
provided the independently characterized external dephasing is
retained unchanged.
Therefore, the Kafri--Taylor--Milburn model does not invalidate the two theorems established here \cite{kafri2013noiseinequalityclassicalforces,Kafri_2014}. 

The status of gravity as a mediator should not be reduced to an isolated math condition. To exclude a concrete classical gravitational description, one must map its state-dependent back-action, its permitted communication records, and its force law onto a definite dynamical channel, then compare that channel with an independently characterized apparatus. The phase-first approximation is compelling in the regime of negligible arm transfer, but it is not a theorem selecting one unique classical communication model. Our second theorem helps bridge that gap by including an active class where information gain and disturbance are explicit. The finite-hopping estimate in Sec.~\ref{sec:robustness_hopping} supplies a complementary statement: passive-model discrimination persists under a sufficiently small, independently controlled dynamical perturbation. Neither result should be converted into a claim about unrestricted classical gravity without specifying the permitted communication and feedback operations. Further enlargements must be similarly precise, rather than obtained by assigning unobserved laboratory noise to the mediator.

Besides, in the all-time separable regime, the total noisy channel
can itself admit a continuous LOCC realization if part
of its dephasing is assigned to measurement back-action.
Our positive gap instead concerns simulations retaining
the full independently calibrated external dephasing
in addition to any mediator-induced disturbance.

We conclude that the disappearance of final-state entanglement need not erase every operational test of gravitational phase dynamics. Within explicitly defined and physically motivated classical mediator classes, conditional-coherence measurements and channel simulation distances remain informative even after entanglement has disappeared. 
Our findings are expected to lay the groundwork for noisy near-term experiments probing the quantum aspects of gravity.

\begin{acknowledgments}
T.F. thanks L. Lami for helpful discussions on an early draft at the conference BYOND IID 2026. T.F. acknowledges support from the CCF--Tencent
Rhino-Bird Open Research Fund (No.~T105-TEG-2024080200001).
V.V. thanks the Gordon and Betty Moore Foundation and the
Conjecture Institute for financial support.
The authors conceived the physical problem, formulated the operational comparison, developed the framework, carried out the derivations, obtained the principal results,
and prepared the original manuscript. AI tools were used during
manuscript preparation to assist with grammar, wording, coding,
algebraic checks, and proof exploration. In particular,
GPT-assisted tools were used for exploratory derivations and
consistency checks in parts of the diagonal-PPT analysis
underlying Theorem~\ref{thm:diagPPT_gap} and
Appendix~\ref{app:diag_ppt_proof}. The authors take full responsibility for the final statements and presentation.
\end{acknowledgments}

\bibliographystyle{apsrev4-2}
\bibliography{ref_qg}

@misc{kafri2013noiseinequalityclassicalforces,
 author = {Kafri, Dvir and Taylor, J. M.},
 title = {A noise inequality for classical forces},
 year = {2013},
 eprint = {1311.4558},
 archivePrefix = {arXiv},
 primaryClass = {quant-ph},
 url = {https://arxiv.org/abs/1311.4558}
}

@article{Kafri_2014,
 author = {Kafri, Dvir and Taylor, J. M. and Milburn, G. J.},
 title = {A classical channel model for gravitational decoherence},
 journal = {New Journal of Physics},
 volume = {16},
 pages = {065020},
 year = {2014},
 doi = {10.1088/1367-2630/16/6/065020}
}

@article{bose_spin_2017,
 author = {Bose, Sougato and Mazumdar, Anupam and Morley, Gavin W. and Ulbricht, Hendrik and Toro\v{s}, Marko and Paternostro, Mauro and Geraci, Andrew A. and Barker, Peter F. and Kim, M. S. and Milburn, Gerard},
 title = {Spin Entanglement Witness for Quantum Gravity},
 journal = {Physical Review Letters},
 volume = {119},
 pages = {240401},
 year = {2017},
 doi = {10.1103/PhysRevLett.119.240401}
}

@article{marletto_gravitationally_2017,
 author = {Marletto, Chiara and Vedral, Vlatko},
 title = {Gravitationally Induced Entanglement between Two Massive Particles is Sufficient Evidence of Quantum Effects in Gravity},
 journal = {Physical Review Letters},
 volume = {119},
 pages = {240402},
 year = {2017},
 doi = {10.1103/PhysRevLett.119.240402}
}

@article{cow1975,
 author = {Colella, R. and Overhauser, A. W. and Werner, S. A.},
 title = {Observation of Gravitationally Induced Quantum Interference},
 journal = {Physical Review Letters}, volume = {34}, pages = {1472--1474},
 year = {1975}, doi = {10.1103/PhysRevLett.34.1472}
}

@article{abeleleeb2012,
 author = {Abele, Hartmut and Leeb, Helmut},
 title = {Gravitation and quantum interference experiments with neutrons},
 journal = {New Journal of Physics}, volume = {14}, pages = {055010},
 year = {2012}, doi = {10.1088/1367-2630/14/5/055010}
}

@article{overstreet2022,
 author = {Overstreet, Chris and Asenbaum, Peter and Curti, Joseph and Kim, Minjeong and Kasevich, Mark A.},
 title = {Observation of a gravitational Aharonov--Bohm effect},
 journal = {Science}, volume = {375}, pages = {226--229}, year = {2022},
 doi = {10.1126/science.abl7152}
}

@article{christodoulou2023,
 author = {Christodoulou, Marios and Di Biagio, Andrea and Aspelmeyer, Markus and Brukner, Caslav and Rovelli, Carlo and Howl, Richard},
 title = {Locally Mediated Entanglement through Gravity from First Principles},
 journal = {Physical Review Letters}, volume = {130}, pages = {100202}, year = {2023},
 doi = {10.1103/PhysRevLett.130.100202}
}

@article{COWPhysRevLett.33.1237,
  title = {Experimental Test of Gravitationally Induced Quantum Interference},
  author = {Overhauser, A. W. and Colella, R.},
  journal = {Phys. Rev. Lett.},
  volume = {33},
  issue = {20},
  pages = {1237--1239},
  numpages = {0},
  year = {1974},
  month = {Nov},
  publisher = {American Physical Society},
  doi = {10.1103/PhysRevLett.33.1237},
  url = {https://link.aps.org/doi/10.1103/PhysRevLett.33.1237}
}

@article{diosi2025no,
  title={No, classical gravity does not entangle quantized matter fields},
  author={Di{\'o}si, Lajos},
  journal={arXiv preprint arXiv:2511.00852},
  year={2025}
}

@article{altamirano2018gravity,
  title={Gravity is not a pairwise local classical channel},
  author={Altamirano, Natacha and Corona-Ugalde, Paulina and Mann, Robert B and Zych, Magdalena},
  journal={Classical and Quantum Gravity},
  volume={35},
  number={14},
  pages={145005},
  year={2018},
  publisher={IOP Publishing}
}

@article{noman2026operational,
  title={Operational Limits of Gravity-Mediated Entanglement},
author = {{Atta ur Rahman} and Noman, Muhammad
          and Zhu, Ying-Jian and Peng, Jia-Xin
          and Haddadi, Saeed},
  journal={arXiv preprint arXiv:2610.22097},
  year={2026}
}

@misc{wang2026recoverabilitygravitationalentanglementdecoherence,
      title={Recoverability of Gravitational Entanglement under Decoherence}, 
      author={Yi-Xuan Wang and Julen S. Pedernales and Martin B. Plenio},
      year={2026},
      eprint={2611.05044},
      archivePrefix={arXiv},
      primaryClass={quant-ph},
      url={https://arxiv.org/abs/2611.05044}, 
}

@article{r8ry-sp35,
  title = {Gravity-induced entanglement is not a theory-independent witness of nonclassicality of the gravitational field},
  author = {Di Biagio, Andrea},
  journal = {Phys. Rev. D},
  volume = {114},
  issue = {8},
  pages = {084050},
  numpages = {12},
  year = {2026},
  month = {Oct},
  publisher = {American Physical Society},
  doi = {10.1103/r8ry-sp35},
  url = {https://link.aps.org/doi/10.1103/r8ry-sp35}
}

@article{plato2023enhanced,
  title={Enhanced gravitational entanglement via modulated optomechanics},
  author={Plato, A Douglas K and R{\"a}tzel, Dennis and Wan, Chuanqi},
  journal={Quantum},
  volume={7},
  pages={1177},
  year={2023},
  publisher={Verein zur F{\"o}rderung des Open Access Publizierens in den Quantenwissenschaften}
}

@misc{feng2026conservationlawsnonclassicalitygravity,
      title={Conservation Laws and the Non-Classicality of Gravity}, 
      author={Tianfeng Feng and Chiara Marletto and Vlatko Vedral},
      year={2026},
      eprint={2311.08971},
      archivePrefix={arXiv},
      primaryClass={quant-ph},
      url={https://arxiv.org/abs/2311.08971}, 
}

@book{dewitt2003global,
  title={The global approach to quantum field theory},
  author={DeWitt, Bryce Seligman},
  volume={114},
  year={2003},
  publisher={Oxford University Press, USA}
}

@article{peres2001hybrid,
  title={Hybrid classical-quantum dynamics},
  author={Peres, Asher and Terno, Daniel R},
  journal={Physical Review A},
  volume={63},
  number={2},
  pages={022101},
  year={2001},
  publisher={APS}
}

@article{eppley1977necessity,
  title={The necessity of quantizing the gravitational field},
  author={Eppley, Kenneth and Hannah, Eric},
  journal={Foundations of Physics},
  volume={7},
  number={1},
  pages={51--68},
  year={1977},
  publisher={Springer}
}

@article{marletto2017we,
  title={Why we need to quantise everything, including gravity},
  author={Marletto, Chiara and Vedral, Vlatko},
  journal={npj Quantum Information},
  volume={3},
  number={1},
  pages={29},
  year={2017},
  publisher={Nature Publishing Group UK, London}
}

@article{CMVVPhysRevD.102.086012,
  title = {Witnessing nonclassicality beyond quantum theory},
  author = {Marletto, Chiara and Vedral, Vlatko},
  journal = {Phys. Rev. D},
  volume = {102},
  issue = {8},
  pages = {086012},
  numpages = {8},
  year = {2020},
  month = {Oct},
  publisher = {American Physical Society},
  doi = {10.1103/PhysRevD.102.086012},
  url = {https://link.aps.org/doi/10.1103/PhysRevD.102.086012}
}

@article{martinmartinez2023,
 author = {Mart{\'i}n-Mart{\'i}nez, Eduardo and Perche, T. Rick},
 title = {What gravity mediated entanglement can really tell us about quantum gravity},
 journal = {Physical Review D}, volume = {108}, pages = {L101702}, year = {2023},
 doi = {10.1103/PhysRevD.108.L101702}
}

@article{marlettoVedral2025RMP,
 author = {Marletto, Chiara and Vedral, Vlatko},
 title = {Quantum-information methods for quantum gravity laboratory-based tests},
 journal = {Reviews of Modern Physics}, volume = {97}, pages = {015006}, year = {2025},
 doi = {10.1103/RevModPhys.97.015006}
}

@article{fengVedral2022,
 author = {Feng, Tianfeng and Vedral, Vlatko},
 title = {Amplification of gravitationally induced entanglement},
 journal = {Physical Review D}, volume = {106}, pages = {066013}, year = {2022},
 doi = {10.1103/PhysRevD.106.066013}
}

@misc{boseRoadmap2025,
 author = {Bose, Sougato and Mazumdar, Anupam and Penrose, Roger and Fuentes, Ivette and Toro{\v s}, Marko and Folman, Ron and Milburn, Gerard J. and Kim, Myungshik and Kent, Adrian and others},
 title = {A Spin-Based Pathway to Testing the Quantum Nature of Gravity},
 year = {2025}, eprint = {2509.01586}, archivePrefix = {arXiv},
 primaryClass = {quant-ph}, url = {https://arxiv.org/abs/2509.01586}
}

@article{azizHowl2025,
 author = {Aziz, Joseph and Howl, Richard},
 title = {Classical theories of gravity produce entanglement},
 journal = {Nature}, volume = {646}, pages = {813--817}, year = {2025},
 doi = {10.1038/s41586-025-09595-7}
}

@article{trilloNavascues2025,
 author = {Trillo, David and Navascu{\'e}s, Miguel},
 title = {Di{\'o}si--Penrose model of classical gravity predicts gravitationally induced entanglement},
 journal = {Physical Review D}, volume = {111}, pages = {L121101}, year = {2025},
 doi = {10.1103/PhysRevD.111.L121101}
}

@article{fengVedralMarletto2026,
 author = {Feng, Tianfeng and Vedral, Vlatko and Marletto, Chiara},
 title = {Collapse-based models for gravity do not violate the entanglement-based witness of nonclassicality},
 journal = {Physical Review D}, volume = {113}, pages = {104055}, year = {2026},
 doi = {10.1103/83rl-nygv}
}

@incollection{feynman1957,
 author = {Feynman, Richard P.},
 title = {Critical Comments},
 booktitle = {The Role of Gravitation in Physics: Report from the 1957 Chapel Hill Conference},
 editor = {DeWitt, Cecile M. and Rickles, Dean},
 publisher = {Edition Open Access}, address = {Berlin}, year = {2011},
 note = {Reprint of proceedings of the 1957 Chapel Hill conference},
 url = {https://edition-open-sources.org/sources/5/5/index.html}
}

@article{pageGeilker1981,
 author = {Page, Don N. and Geilker, C. D.},
 title = {Indirect Evidence for Quantum Gravity},
 journal = {Physical Review Letters}, volume = {47}, pages = {979--982}, year = {1981},
 doi = {10.1103/PhysRevLett.47.979}
}

@article{lami2024,
 author = {Lami, Ludovico and Pedernales, Julen S. and Plenio, Martin B.},
 title = {Testing the Quantumness of Gravity without Entanglement},
 journal = {Physical Review X}, volume = {14}, pages = {021022}, year = {2024},
 doi = {10.1103/PhysRevX.14.021022}
}

@article{hornberger2003,
 author = {Hornberger, Klaus and Uttenthaler, Stefan and Brezger, Bjorn and Hackermuller, Lucia and Arndt, Markus and Zeilinger, Anton},
 title = {Collisional Decoherence Observed in Matter Wave Interferometry},
 journal = {Physical Review Letters}, volume = {90}, pages = {160401}, year = {2003},
 doi = {10.1103/PhysRevLett.90.160401}
}

@article{hackermueller2004,
 author = {Hackermuller, Lucia and Hornberger, Klaus and Brezger, Bjorn and Zeilinger, Anton and Arndt, Markus},
 title = {Decoherence of Matter Waves by Thermal Emission of Radiation},
 journal = {Nature}, volume = {427}, pages = {711--714}, year = {2004},
 doi = {10.1038/nature02276}
}

@article{hornbergerSipe2003,
 author = {Hornberger, Klaus and Sipe, John E.}, title = {Collisional Decoherence Reexamined},
 journal = {Physical Review A}, volume = {68}, pages = {012105}, year = {2003},
 doi = {10.1103/PhysRevA.68.012105}
}

@article{hornbergerSipeArndt2004,
 author = {Hornberger, Klaus and Sipe, John E. and Arndt, Markus},
 title = {Theory of Decoherence in a Matter Wave Talbot--Lau Interferometer},
 journal = {Physical Review A}, volume = {70}, pages = {053608}, year = {2004},
 doi = {10.1103/PhysRevA.70.053608}
}

@article{marshman2020,
 author = {Marshman, Ryan J. and Mazumdar, Anupam and Bose, Sougato},
 title = {Locality and Entanglement in Table-Top Testing of the Quantum Nature of Linearized Gravity},
 journal = {Physical Review A}, volume = {101}, pages = {052110}, year = {2020},
 doi = {10.1103/PhysRevA.101.052110}
}

@article{vandeKamp2020,
 author = {van de Kamp, Thomas W. and Marshman, Ryan J. and Bose, Sougato and Mazumdar, Anupam},
 title = {Quantum Gravity Witness via Entanglement of Masses: Casimir Screening},
 journal = {Physical Review A}, volume = {102}, pages = {062807}, year = {2020},
 doi = {10.1103/PhysRevA.102.062807}
}

@article{krisnanda2017,
 author = {Krisnanda, Tanjung and Zuppardo, Margherita and Paternostro, Mauro and Paterek, Tomasz},
 title = {Revealing Nonclassicality of Inaccessible Objects},
 journal = {Physical Review Letters}, volume = {119}, pages = {120402}, year = {2017},
 doi = {10.1103/PhysRevLett.119.120402}
}

@article{galley2022,
 author = {Galley, Thomas D. and Giacomini, Flaminia and Selby, John H.},
 title = {A No-Go Theorem on the Nature of the Gravitational Field beyond Quantum Theory},
 journal = {Quantum}, volume = {6}, pages = {779}, year = {2022}
}

@article{danielson2022,
 author = {Danielson, Daine L. and Satishchandran, Gautam and Wald, Robert M.},
 title = {Gravitationally Mediated Entanglement: Newtonian Field versus Gravitons},
 journal = {Physical Review D}, volume = {105}, pages = {086001}, year = {2022},
 doi = {10.1103/PhysRevD.105.086001}
}

@article{carney2022,
 author = {Carney, Daniel}, title = {Newton, Entanglement, and the Graviton},
 journal = {Physical Review D}, volume = {105}, pages = {024029}, year = {2022},
 doi = {10.1103/PhysRevD.105.024029}
}

@article{fragkos2022,
 author = {Fragkos, Vasilis and Kopp, Michael and Pikovski, Igor},
 title = {On Inference of Quantization from Gravitationally Induced Entanglement},
 journal = {AVS Quantum Science}, volume = {4}, pages = {045601}, year = {2022}
}

@article{fein2019,
 author = {Fein, Yaakov Y. and Geyer, Philipp and Zwick, Patrick and Kialka, Filip and Pedalino, Sebastian and Mayor, Marcel and Gerlich, Stefan and Arndt, Markus},
 title = {Quantum Superposition of Molecules beyond 25 {kDa}},
 journal = {Nature Physics}, volume = {15}, pages = {1242--1245}, year = {2019}
}

@article{westphal2021,
 author = {Westphal, Tobias and Hepach, Hans and Pfaff, Jeremias and Aspelmeyer, Markus},
 title = {Measurement of Gravitational Coupling between Millimetre-Sized Masses},
 journal = {Nature}, volume = {591}, pages = {225--228}, year = {2021}
}

@article{pedernales2020,
 author = {Pedernales, Julen S. and Morley, Gavin W. and Plenio, Martin B.},
 title = {Motional Dynamical Decoupling for Interferometry with Macroscopic Particles},
 journal = {Physical Review Letters}, volume = {125}, pages = {023602}, year = {2020},
 doi = {10.1103/PhysRevLett.125.023602}
}

@article{pedernales2022,
 author = {Pedernales, Julen S. and Streltsov, Kirill and Plenio, Martin B.},
 title = {Enhancing Gravitational Interaction between Quantum Systems by a Massive Mediator},
 journal = {Physical Review Letters}, volume = {128}, pages = {110401}, year = {2022},
 doi = {10.1103/PhysRevLett.128.110401}
}

@article{cosco2021,
 author = {Cosco, Filippo and Pedernales, Julen S. and Plenio, Martin B.},
 title = {Enhanced Force Sensitivity and Entanglement in Periodically Driven Optomechanics},
 journal = {Physical Review A}, volume = {103}, pages = {L061501}, year = {2021},
 doi = {10.1103/PhysRevA.103.L061501}
}

@article{weiss2021,
 author = {Weiss, Thomas and Roda-Llordes, Marc and Torrontegui, Erik and Aspelmeyer, Markus and Romero-Isart, Oriol},
 title = {Large Quantum Delocalization of a Levitated Nanoparticle Using Optimal Control: Applications for Force Sensing and Entangling via Weak Forces},
 journal = {Physical Review Letters}, volume = {127}, pages = {023601}, year = {2021},
 doi = {10.1103/PhysRevLett.127.023601}
}

@article{krisnanda2020,
 author = {Krisnanda, Tanjung and Tham, G. Y. and Paternostro, Mauro and Paterek, Tomasz},
 title = {Observable Quantum Entanglement Due to Gravity},
 journal = {npj Quantum Information}, volume = {6}, pages = {12}, year = {2020}
}

@article{miao2020,
 author = {Miao, Haixing and Martynov, Denis and Yang, Huan and Datta, Animesh},
 title = {Quantum Correlations of Light Mediated by Gravity},
 journal = {Physical Review A}, volume = {101}, pages = {063804}, year = {2020},
 doi = {10.1103/PhysRevA.101.063804}
}

@article{chevalier2020,
 author = {Chevalier, H. and Paige, A. J. and Kim, M. S.},
 title = {Witnessing the Nonclassical Nature of Gravity in the Presence of Unknown Interactions},
 journal = {Physical Review A}, volume = {102}, pages = {022428}, year = {2020},
 doi = {10.1103/PhysRevA.102.022428}
}

@article{howl2021,
 author = {Howl, Richard and Vedral, Vlatko and Naik, D. and Christodoulou, Marios and Rovelli, Carlo and Iyer, A.},
 title = {Non-Gaussianity as a Signature of a Quantum Theory of Gravity},
 journal = {PRX Quantum}, volume = {2}, pages = {010325}, year = {2021}
}

@article{carlesso2019,
 author = {Carlesso, Matteo and Bassi, Angelo and Paternostro, Mauro and Ulbricht, Hendrik},
 title = {Testing the Gravitational Field Generated by a Quantum Superposition},
 journal = {New Journal of Physics}, volume = {21}, pages = {093052}, year = {2019}
}

@article{haine2021,
 author = {Haine, Simon A.}, title = {Searching for Signatures of Quantum Gravity in Quantum Gases},
 journal = {New Journal of Physics}, volume = {23}, pages = {033020}, year = {2021}
}

@article{tilloyDiosi2016,
 author = {Tilloy, Antoine and Diosi, Lajos},
 title = {Sourcing Semiclassical Gravity from Spontaneously Localized Quantum Matter},
 journal = {Physical Review D}, volume = {93}, pages = {024026}, year = {2016},
 doi = {10.1103/PhysRevD.93.024026}
}

@article{oppenheim2023,
 author = {Oppenheim, Jonathan}, title = {A Postquantum Theory of Classical Gravity?},
 journal = {Physical Review X}, volume = {13}, pages = {041040}, year = {2023},
 doi = {10.1103/PhysRevX.13.041040}
}

@article{oppenheimDiffusion2023,
 author = {Oppenheim, Jonathan and Sparaciari, Carlo and Soda, Barbara and Weller-Davies, Zachary},
 title = {Gravitationally Induced Decoherence vs Space-Time Diffusion: Testing the Quantum Nature of Gravity},
 journal = {Nature Communications}, volume = {14}, pages = {7910}, year = {2023}
}

@article{kafri2015,
 author = {Kafri, Dvir and Milburn, G. J. and Taylor, J. M.},
 title = {Bounds on Quantum Communication via Newtonian Gravity},
 journal = {New Journal of Physics}, volume = {17}, pages = {015006}, year = {2015}
}

@article{pfister2016,
 author = {Pfister, Corsin and Kaniewski, Jedrzej and Tomamichel, Marco and Mantri, Atul and Schmucker, Rainer and McMahon, Nathan and Milburn, Gerard and Wehner, Stephanie},
 title = {A Universal Test for Gravitational Decoherence},
 journal = {Nature Communications}, volume = {7}, pages = {13022}, year = {2016}
}

@article{carneyStampTaylor2019,
 author = {Carney, Daniel and Stamp, P. C. E. and Taylor, J. M.},
 title = {Tabletop Experiments for Quantum Gravity: A User's Manual},
 journal = {Classical and Quantum Gravity}, volume = {36}, pages = {034001}, year = {2019}
}

@article{peres1996,
 author = {Peres, Asher}, title = {Separability Criterion for Density Matrices},
 journal = {Physical Review Letters}, volume = {77}, pages = {1413--1415}, year = {1996},
 doi = {10.1103/PhysRevLett.77.1413}
}

@article{horodeckiPPT1996,
 author = {Horodecki, Michal and Horodecki, Pawel and Horodecki, Ryszard},
 title = {Separability of Mixed States: Necessary and Sufficient Conditions},
 journal = {Physics Letters A}, volume = {223}, pages = {1--8}, year = {1996}
}

@article{penrose1996,
 author = {Penrose, Roger}, title = {On Gravity's Role in Quantum State Reduction},
 journal = {General Relativity and Gravitation}, volume = {28}, pages = {581--600}, year = {1996}
}

@article{diosi1987,
 author = {Diosi, Lajos},
 title = {A Universal Master Equation for the Gravitational Violation of Quantum Mechanics},
 journal = {Physics Letters A}, volume = {120}, pages = {377--381}, year = {1987}
}

@article{streltsov2022,
 author = {Streltsov, Kirill and Pedernales, Julen S. and Plenio, Martin B.},
 title = {On the Significance of Interferometric Revivals for the Fundamental Description of Gravity},
 journal = {Universe}, volume = {8}, pages = {58}, year = {2022}
}

@article{spaventa2023,
 author = {Spaventa, Giorgio and Lami, Ludovico and Plenio, Martin B.},
 title = {On Tests of the Quantum Nature of Gravitational Interactions in Presence of Non-Linear Corrections to Quantum Mechanics},
 journal = {Quantum}, volume = {7}, pages = {1157}, year = {2023}
}

@article{lindnerPeres2005,
 author = {Lindner, Netanel H. and Peres, Asher},
 title = {Testing Quantum Superpositions of the Gravitational Field with Bose--Einstein Condensates},
 journal = {Physical Review A}, volume = {71}, pages = {024101}, year = {2005},
 doi = {10.1103/PhysRevA.71.024101}
}

@article{sidajaya2022,
 author = {Sidajaya, Putra and Cong, Wan and Scarani, Valerio},
 title = {Possibility of Detecting the Gravity of an Object Frozen in a Spatial Superposition by the Zeno Effect},
 journal = {Physical Review A}, volume = {106}, pages = {042217}, year = {2022},
 doi = {10.1103/PhysRevA.106.042217}
}

@article{christodoulou2023Degrees,
 author = {Christodoulou, Marios and Di Biagio, Andrea and Howl, Richard and Rovelli, Carlo},
 title = {Gravity Entanglement, Quantum Reference Systems, Degrees of Freedom},
 journal = {Classical and Quantum Gravity}, volume = {40}, pages = {047001}, year = {2023}
}

@misc{boseWhitepaper2026,
  author = {Bose, Sougato and Mazumdar, Anupam and Toro{\v{s}}, Marko and Zhou, Tian and Adach, Tadeusz and others},
  title = {White paper: 1--10 {Hz} matter-wave interferometer to test the spin entanglement witness for quantum gravity},
  year = {2026},
  eprint = {2610.07422},
  archivePrefix = {arXiv},
  primaryClass = {quant-ph},
  url = {https://arxiv.org/abs/2610.07422}
}

\clearpage
\onecolumngrid
\appendix

\section{Exact integration of the master equation with local dephasing}
\label{app:exact_dynamics}

The total Lindbladian governing the bipartite open system separates into the Newtonian phase dynamics and the local environmental noise,
\begin{equation}
\mathcal L[\rho]
=
-\frac{i}{\hbar}[H_g,\rho]
+
\gamma\left(Z_A\rho Z_A-\rho\right)
+
\gamma\left(Z_B\rho Z_B-\rho\right),
\label{eq:app_lindbladian}
\end{equation}
where
$
H_g=-\hbar\Gamma_g Z_AZ_B.
$
Both the Hamiltonian and the dephasing operators are diagonal in the computational path basis
\(\{|00\rangle,|01\rangle,|10\rangle,|11\rangle\}\). Hence each matrix element evolves independently.

Let \(z_0=+1\) and \(z_1=-1\). For a basis state \(|ij\rangle\), the Hamiltonian eigenvalue is
$
E_{ij}
=
-\hbar\Gamma_g z_i z_j .
$
Therefore a general density-matrix element \(\rho_{ij,kl}=\langle ij|\rho|kl\rangle\) obeys
\begin{equation}
\dot\rho_{ij,kl}(t)
=
\left[
i\Gamma_g(z_i z_j-z_k z_l)
-
\gamma(1-z_i z_k)
-
\gamma(1-z_j z_l)
\right]
\rho_{ij,kl}(t).
\label{eq:app_element_evolution}
\end{equation}
The two dephasing terms count whether the \(A\) and \(B\) path labels differ between bra and ket. A single-path coherence therefore decays with rate \(2\gamma\), while a two-path coherence decays with rate \(4\gamma\).

For example, for the single-excitation coherence \(\rho_{00,01}\), one has \(z_0z_0=+1\) and \(z_0z_1=-1\). Thus
$
\dot\rho_{00,01}(t)
=
\left(
i2\Gamma_g-2\gamma
\right)
\rho_{00,01}(t),
$
and hence
$
\rho_{00,01}(\tau)
=
e^{-2\gamma\tau}e^{i2\Gamma_g\tau}
\rho_{00,01}(0).
$
With the definitions
$
\Theta=\Gamma_g\tau
$ and $
r=e^{-2\gamma\tau},
$
this factor is \(r e^{i2\Theta}\).

Applying Eq.~\eqref{eq:app_element_evolution} to all matrix elements of the input
\(\rho_1=|++\rangle\langle++|\) gives
\begin{equation}
\sigma_1
=
\frac14
\begin{pmatrix}
1 & r e^{i2\Theta} & r e^{i2\Theta} & r^2\\
r e^{-i2\Theta} & 1 & r^2 & r e^{-i2\Theta}\\
r e^{-i2\Theta} & r^2 & 1 & r e^{-i2\Theta}\\
r^2 & r e^{i2\Theta} & r e^{i2\Theta} & 1
\end{pmatrix},
\end{equation}
which is Eq.~\eqref{eq:sigma_plus} of the main text.

Since the Hamiltonian generator and the dephasing generator are diagonal in the same basis, they commute as superoperators. Therefore the finite-time channel factorizes exactly as
$
\mathcal E_Q(\tau)
=
\mathcal D_r\circ\mathcal U_\Theta,
$
where
$
\mathcal U_\Theta(\rho)=e^{i\Theta Z_AZ_B}\rho e^{-i\Theta Z_AZ_B}.
$

\section{Bell-basis spectrum and PPT threshold}
\label{app:eigenvalues}

We aim to find the exact spectral decomposition of the final decohered quantum state $\sigma_1$. We project $\sigma_1$ into the Bell basis: $\ket{\Phi^\pm} = \frac{1}{\sqrt{2}}(\ket{00} \pm \ket{11})$ and $\ket{\Psi^\pm} = \frac{1}{\sqrt{2}}(\ket{01} \pm \ket{10})$.
Direct evaluation gives
$
    \bra{\Phi^\pm} \sigma_1 \ket{\Phi^\pm} = \frac{1 \pm r^2}{4}$, $
    \bra{\Psi^\pm} \sigma_1 \ket{\Psi^\pm} = \frac{1 \pm r^2}{4},
$
and
$
    \bra{\Phi^+} \sigma_1 \ket{\Psi^+}  = \frac{r}{2} e^{i 2\Theta}$, $
    \bra{\Phi^-} \sigma_1 \ket{\Psi^-} = 0. 
$
Thus, the density matrix in the ordered basis $\{\ket{\Phi^+}, \ket{\Psi^+}, \ket{\Phi^-}, \ket{\Psi^-}\}$ is strictly block-diagonal:
\begin{equation}
    \sigma_1^{(\text{Bell})} = \begin{pmatrix}
        \frac{1+r^2}{4} & \frac{r}{2} e^{i 2\Theta} & 0 & 0 \\
        \frac{r}{2} e^{-i 2\Theta} & \frac{1+r^2}{4} & 0 & 0 \\
        0 & 0 & \frac{1-r^2}{4} & 0 \\
        0 & 0 & 0 & \frac{1-r^2}{4}
    \end{pmatrix}.
\end{equation}
The lower $2 \times 2$ block provides two degenerate eigenvalues $\lambda_3 = \lambda_4 = \frac{1-r^2}{4}$.
The upper $2 \times 2$ block can be rewritten as $\frac{1+r^2}{4} I + \frac{r}{2} (\cos(2\Theta) \sigma_x - \sin(2\Theta) \sigma_y)$. Its eigenvalues are trivially found by adding/subtracting the norm of the off-diagonal vector $\frac{r}{2}$, yielding $\lambda_{1,2} = \frac{(1 \pm r)^2}{4}$.

Applying the Peres-Horodecki criterion involves performing a partial transpose $T_B$. 

The spectrum of the partially transposed state is
\begin{equation}
\operatorname{spec}(\sigma_1^{T_B})
=
\left\{
\frac{1+r^2+2r\cos 2\Theta}{4},
\frac{1+r^2-2r\cos 2\Theta}{4},
\frac{1-r^2+2r\sin 2\Theta}{4},
\frac{1-r^2-2r\sin 2\Theta}{4}
\right\},
\end{equation}
up to the replacement \(\sin 2\Theta\mapsto |\sin 2\Theta|\) when identifying the smallest eigenvalue. 
Since the \(\cos 2\Theta\) branch is always
nonnegative, the only eigenvalue branch
that can become negative is
\begin{equation}
\lambda_{\mathrm{crit}}
=
\frac14\left(1-r^2-2r|\sin2\Theta|\right).
\end{equation}
This critical branch need not be the smallest
eigenvalue when the state is PPT; its sign
nonetheless gives the exact entanglement criterion.

\section{PPT lower bound for a pure gravitational-phase output}
\label{app:ppt_lower_bound}

In this appendix, we prove a lower bound on the trace distance between an
entangled two-qubit pure state and the set of PPT states. This elementary bound also supplies the PPT-state comparison used in the noiseless channel and Choi analysis.

Let \(|\psi\rangle\) be a two-qubit pure state with Schmidt decomposition
$
|\psi\rangle
=
s_1|00\rangle+s_2|11\rangle,
$$
s_1,s_2\geq0,
$ and $
s_1^2+s_2^2=1.
$
Taking the partial transpose on subsystem \(B\), we obtain
\[
\rho_\psi^{T_B}
=
s_1^2|00\rangle\langle00|
+
s_2^2|11\rangle\langle11|
+
s_1s_2
\left(
|01\rangle\langle10|
+
|10\rangle\langle01|
\right).
\]
Its spectrum is
$
\operatorname{spec}(\rho_\psi^{T_B})
=
\{s_1^2,\ s_2^2,\ s_1s_2,\ -s_1s_2\}.
$
The negative eigenvector is
$
|\eta_-\rangle
=
\frac{1}{\sqrt2}\left(|01\rangle-|10\rangle\right),
$
with eigenvalue
$
\rho_\psi^{T_B}|\eta_-\rangle
=
-s_1s_2|\eta_-\rangle.
$
Define
$
\Pi_- = |\eta_-\rangle\langle\eta_-|.
$
For any PPT state \(\sigma\), we have
$
\sigma^{T_B}\succeq0,
$
and hence
$
\operatorname{Tr}[\Pi_-\sigma^{T_B}]\geq0.
$
On the other hand,
$
\operatorname{Tr}[\Pi_-\rho_\psi^{T_B}]
=
-s_1s_2.
$
Therefore
\[
\operatorname{Tr}\left[
\Pi_-
\left(
\sigma^{T_B}-\rho_\psi^{T_B}
\right)
\right]
\geq
s_1s_2.
\]
Note that
$
\operatorname{Tr}\left[
\Pi_-
\left(
\sigma^{T_B}-\rho_\psi^{T_B}
\right)
\right]
=
\operatorname{Tr}\left[
\Pi_-^{T_B}
\left(
\sigma-\rho_\psi
\right)
\right], 
$
and  
$
\Pi_-^{T_B}
$ is a Hermitian operator.
It is easy to verify that the eigenvalues of \(\Pi_-^{T_B}\) are
$
\left\{
\frac12,\frac12,\frac12,-\frac12
\right\},
$
so
$
\|\Pi_-^{T_B}\|_\infty=\frac12.
$
By Hölder inequality,
\[
\left|
\operatorname{Tr}\left[
\Pi_-^{T_B}(\sigma-\rho_\psi)
\right]
\right|
\leq
\|\Pi_-^{T_B}\|_\infty
\|\sigma-\rho_\psi\|_1
=
\frac12
\|\sigma-\rho_\psi\|_1.
\]
Combining this with the previous inequality gives
\[
s_1s_2 \le\operatorname{Tr}\left[
\Pi_-^{T_B}
\left(
\sigma-\rho_\psi
\right)
\right]
\leq
\frac12
\|\sigma-\rho_\psi\|_1.
\]
Thus, for every PPT state \(\sigma\),
$
\frac12
\|\sigma-\rho_\psi\|_1
\geq
s_1s_2.
$
Taking the infimum over all PPT states yields
\[
{
\inf_{\sigma\in{\rm PPT}}
\frac12
\left\|
|\psi\rangle\langle\psi|-\sigma
\right\|_1
\geq
s_1s_2.
}
\]

For the state generated by the noiseless gravitational unitary
\(U_\Theta=e^{i\Theta Z_AZ_B}\) from the input \(|++\rangle\),
\[
|\psi_\Theta\rangle
=
\frac12
\left(
e^{i\Theta}|00\rangle
+
e^{-i\Theta}|01\rangle
+
e^{-i\Theta}|10\rangle
+
e^{i\Theta}|11\rangle
\right),
\]
the reduced density matrix has eigenvalues
$
\lambda_\pm
=
\frac{1\pm \cos2\Theta}{2}.
$
Therefore the product of the Schmidt coefficients is
$
s_1s_2
=
\sqrt{\lambda_+\lambda_-}
=
\frac{|\sin2\Theta|}{2}.
$
Hence
\[
{
\inf_{\sigma\in{\rm PPT}}
\frac12
\left\|
|\psi_\Theta\rangle\langle\psi_\Theta|
-
\sigma
\right\|_1
\geq
\frac{|\sin2\Theta|}{2}.
}
\]
This proves the bound used in the main text.

\section{Saturation of the LP gap and terminal-time classical dynamics}
\label{app:LP_proof}
\subsection{Conditional coherence witness and exact trace norm}
Let $F=Z_AZ_B$ and $\rho_1=|++\rangle\langle++|$. The dephasing map multiplies matrix elements differing by one path bit by $r$ and those differing by two path bits by $r^2$. Every product-diagonal unitary $U_A(\lambda)\otimes U_B(\lambda)$ multiplies the two matrix elements $\rho_{00,01}$ and $\rho_{10,11}$ by the same factor $e^{i(\beta_0-\beta_1)}$, so for every mixture of these channels,
\begin{equation}
(\tilde\sigma_1)_{00,01}=(\tilde\sigma_1)_{10,11}.
\label{eq:app_LP_equal}
\end{equation}
By direct evaluation of the target state,
\begin{equation}
\operatorname{Tr}[-Z_AY_B\,\sigma_1]=r\sin2\Theta,
\quad\operatorname{Tr}[-Z_AY_B\,\tilde\sigma_1]=0.
\end{equation}
Because $\|-Z_AY_B\|_\infty=1$, trace-distance duality establishes
\begin{equation}
\inf_{\mathcal C_{\rm LP}}D(\sigma_1,\Dd_r\circ\mathcal C_{\rm LP}(\rho_1))
\ge\frac r2|\sin2\Theta|.
\label{eq:app_LP_lower}
\end{equation}

Choose the shared-randomness mixture $\mathcal C_*=\cos^2\Theta\,\mathcal I+\sin^2\Theta\,\mathcal Z_{AB}$, where $\mathcal Z_{AB}(\rho)=F\rho F$. This lies in LP for every $\Theta$. Expanding $U_\Theta=\cos\Theta I+i\sin\Theta F$ gives
\begin{align}
\cU_\Theta(\rho_1)-\mathcal C_*(\rho_1)
&=i\cos\Theta\sin\Theta[F,\rho_1]\\
&=-\frac{\sin2\Theta}{4}(Y_AZ_B+Z_AY_B).
\end{align}
Applying local dephasing yields Eq.~\eqref{eq:LP_optimal_difference}, since both $Y_AZ_B$ and $Z_AY_B$ contain one transverse Pauli operator. The two operators commute, square to $I$, and have product $X_AX_B$. Hence $(Y_AZ_B+Z_AY_B)^2=2(I+X_AX_B)$, with eigenvalues $4,4,0,0$. Consequently the difference spectrum is $\{\pm r\sin2\Theta/2,0,0\}$ and its trace norm saturates Eq.~\eqref{eq:app_LP_lower}. This proves Theorem~\ref{thm:LP_phase_gap} without invoking numerics.

\subsection{Interleaving dephasing with arbitrary LP steps}
For a finite subdivision of the duration $\tau$, let $\Dd_{r_j}$ represent external local dephasing over interval $j$ and $\mathcal C_j$ any LP channel. Since $\Dd_{r_j}$ is diagonal in the path operator basis, it commutes with every product-diagonal unitary and hence with every $\mathcal C_j$. The class LP is closed under composition: the product of two local diagonal unitaries remains a local diagonal unitary, and the joint classical random variables can be incorporated into one shared random variable. It follows that
\begin{equation}
(\Dd_{r_N}\circ\mathcal C_N)\circ\cdots\circ
(\Dd_{r_1}\circ\mathcal C_1)=
\Dd_{\prod_j r_j}\circ (\mathcal C_N\circ\cdots\circ\mathcal C_1).
\label{eq:interleaved_LP_factorization}
\end{equation}
Thus the finite-time theorem applies to all such interleavings. Optimizing over an arbitrary terminal LP map is at least as permissive as optimizing over a specified finite-step stochastic process, and the lower bound holds for both. Whether the bound is attained inside a more restricted time-homogeneous family requires an explicit construction, given next.

\subsection{Time-homogeneous realization at a prescribed terminal time}
Let $\mathcal Z_{AB}(\rho)=F\rho F$ and define the generator
\begin{equation}
\mathcal L_*(\rho)=\kappa(\mathcal Z_{AB}-\mathcal I)(\rho),\qquad\kappa\ge0.
\end{equation}
This is the generator of a classical Poisson process of correlated local phase flips: at each jump, Alice and Bob separately apply their respective $Z$ operators, and only the shared classical jump record connects them. Since $\mathcal Z_{AB}^2=\mathcal I$, its semigroup is
\begin{equation}
e^{t\mathcal L_*}=\frac{1+e^{-2\kappa t}}2\,\mathcal I
+\frac{1-e^{-2\kappa t}}2\,\mathcal Z_{AB}.
\label{eq:LP_poisson_semigroup}
\end{equation}
When $\cos2\Theta>0$, choosing $\kappa=-(2\tau)^{-1}\ln(\cos2\Theta)$ gives $e^{\tau\mathcal L_*}=\mathcal C_*$ and therefore exactly saturates the finite-time gap, even when local external dephasing acts continuously and concurrently. For $\cos2\Theta=0$, the finite rate construction only approaches the optimum in the limit $\kappa\tau\to\infty$. For $\cos2\Theta<0$, this particular purely dissipative semigroup does not represent $\mathcal C_*$, although a terminal-time random LP channel still does. The rate $\kappa$ is chosen for the specified terminal point $\tau$, and it is not a single $\tau$-independent optimizer for the entire family $\{\mathcal U_{\Gamma_g t}\}_{t\ge0}$.


\section{Noiseless Choi equivalence of PPT and local-phase models}
\label{app:choi_equivalence}

In this appendix, we prove the noiseless Choi-distance equivalence stated in the
main text. We use the normalized Choi state
$
    \omega_\Phi
    =
    (\Phi\otimes{\rm id})(|\Phi_4\rangle\langle\Phi_4|)
    =
    \frac{1}{4}J_\Phi ,
$
where
$
    |\Phi_4\rangle
    =
    \frac{1}{2}
    \sum_{\mu=0}^{3}
    |\mu\rangle_{A_oB_o}
    |\mu\rangle_{A_iB_i}
$
is the maximally entangled state between the two-qubit output and input
systems. Thus the Choi trace distance used here is
$
    D_{\rm Choi}(\Phi,\Psi)
    =
    \frac12
    \|\omega_\Phi-\omega_\Psi\|_1 .
$
Equivalently, in terms of unnormalized Choi matrices,
\begin{equation}
    D_{\rm Choi}(\Phi,\Psi)
    =
    \frac18
    \|J_\Phi-J_\Psi\|_1 .
\end{equation}

Let
$
    F=Z_A\otimes Z_B .
$
The ideal gravitational unitary is
$
    U_\Theta
    =
    e^{i\Theta F}
    =
    \cos\Theta\, I
    +
    i\sin\Theta\, F .
$
For a unitary channel \({\cal U}_\Theta(\rho)=U_\Theta\rho U_\Theta^\dagger\),
the normalized Choi state is pure:
$
    \omega_{{\cal U}_\Theta}
    =
    |U_\Theta\rangle\rangle\langle\langle U_\Theta| ,
$
where
$
    |U_\Theta\rangle\rangle
    =
    (U_\Theta\otimes I)|\Phi_4\rangle .
$
With the same normalized convention, define
$
    |I\rangle\rangle
    =
    (I\otimes I)|\Phi_4\rangle,
 $ $
    |F\rangle\rangle
    =
    (F\otimes I)|\Phi_4\rangle .
$
Since \(\operatorname{Tr}F=0\) and \(F^2=I\), these two Choi vectors are
orthonormal:
$
    \langle\langle I|F\rangle\rangle
    =
    \frac{1}{4}\operatorname{Tr}F
    =
    0,
$ $
    \langle\langle I|I\rangle\rangle
    =
    \langle\langle F|F\rangle\rangle
    =
    1 .
$
Therefore, we have
\begin{equation}
    |U_\Theta\rangle\rangle
    =
    \cos\Theta\, |I\rangle\rangle
    +
    i\sin\Theta\, |F\rangle\rangle .
\end{equation}

We now view the Choi state as a bipartite state across the laboratory cut
$
    A_oA_i\,|\,B_oB_i .
$
The maximally entangled state factorizes across Alice and Bob as
$
    |\Phi_4\rangle_{A_oB_o:A_iB_i}
    =
    |\Phi_2\rangle_{A_oA_i}
    \otimes
    |\Phi_2\rangle_{B_oB_i},
$
up to the fixed permutation that groups Alice's and Bob's systems. Hence
$
    |I\rangle\rangle
    =
    |I_A\rangle\rangle\otimes |I_B\rangle\rangle ,
$ $
    |F\rangle\rangle
    =
    |Z_A\rangle\rangle\otimes |Z_B\rangle\rangle ,
$
where
$
    |I_A\rangle\rangle
    =
    (I_A\otimes I_{A_i})|\Phi_2\rangle_{A_oA_i}
$ and $
    |Z_A\rangle\rangle
    =
    (Z_A\otimes I_{A_i})|\Phi_2\rangle_{A_oA_i},
$
and similarly for Bob. Since
$
    \langle\langle I_A|Z_A\rangle\rangle
    =
    \frac12\operatorname{Tr}Z_A
    =
    0,
$ $
    \langle\langle I_B|Z_B\rangle\rangle
    =
    0,
$
the decomposition
\begin{equation}
    |U_\Theta\rangle\rangle
    =
    \cos\Theta\,
    |I_A\rangle\rangle |I_B\rangle\rangle
    +
    i\sin\Theta\,
    |Z_A\rangle\rangle |Z_B\rangle\rangle
\end{equation}
is already a Schmidt decomposition across \(A_oA_i\,|\,B_oB_i\), up to
irrelevant local phases. Thus the two nonzero Schmidt coefficients of the
ideal Choi state are
$
    s_1=|\cos\Theta|
$ and $
    s_2=|\sin\Theta|.
$

We next prove a lower bound against the full PPT-Choi relaxation. Let
\(\Lambda\) be any PPT-Choi channel, meaning
$
    J_\Lambda\ge0,
$ $
    \operatorname{Tr}_{A_oB_o}J_\Lambda=I_{A_iB_i}, 
$ and $
    J_\Lambda^{T_{B_oB_i}}\ge0 .
$
Then \(\omega_\Lambda=J_\Lambda/4\) is a normalized PPT state across the cut
\(A_oA_i\,|\,B_oB_i\). Therefore the distance from
\(\omega_{{\cal U}_\Theta}\) to the PPT-Choi set is lower bounded by its
distance to the full PPT state set.

For completeness, we recall the elementary Schmidt-rank-two bound. Let
\begin{equation}
        |\psi\rangle
    =
    s_1|a_0\rangle|b_0\rangle
    +
    s_2|a_1\rangle|b_1\rangle
\end{equation}
be a bipartite pure state with two nonzero Schmidt coefficients. The partial
transpose of \(|\psi\rangle\langle\psi|\) has a negative eigenvalue
\(-s_1s_2\), with eigenvector
$
    |\eta_-\rangle
    =
    \frac{1}{\sqrt2}
    \left(
    |a_0\rangle|b_1\rangle
    -
    |a_1\rangle|b_0\rangle
    \right).
$
Let \(\Pi_-=|\eta_-\rangle\langle\eta_-|\). For any PPT state \(\sigma\),
$
    \operatorname{Tr}[\Pi_- \sigma^{T_B}]\ge0
$ and $
    \operatorname{Tr}[\Pi_- (|\psi\rangle\langle\psi|)^{T_B}]
    =
    -s_1s_2 .
$
Hence, we have
$
    \operatorname{Tr}
    \left[
    \Pi_-^{T_B}
    \left(
    \sigma-|\psi\rangle\langle\psi|
    \right)
    \right]
    \ge
    s_1s_2 .
$
Moreover,
$
    \|\Pi_-^{T_B}\|_\infty=\frac12 ,
$
resulting
\begin{equation}
    s_1s_2
    \le
    \left|
    \operatorname{Tr}
    \left[
    \Pi_-^{T_B}
    \left(
    \sigma-|\psi\rangle\langle\psi|
    \right)
    \right]
    \right|
    \le
    \frac12
    \left\|
    \sigma-|\psi\rangle\langle\psi|
    \right\|_1 .
\end{equation}
Thus, one has
\begin{equation}
    \inf_{\sigma\in{\rm PPT}}
    \frac12
    \left\|
    \sigma-|\psi\rangle\langle\psi|
    \right\|_1
    \ge
    s_1s_2 .
\end{equation}
Applying this to the ideal Choi state gives
\begin{equation}
    \inf_{\Lambda\in{\rm PPT\text{-}Choi}}
    D_{\rm Choi}({\cal U}_\Theta,\Lambda)
    \ge
    |\cos\Theta\,\sin\Theta|
    =
    \frac{|\sin2\Theta|}{2}.
\end{equation}

It remains to show that this lower bound is achievable by a path-local phase
channel. Consider
$
    \Lambda_*
    =
    \cos^2\Theta\,{\cal I}
    +
    \sin^2\Theta\,{\cal Z}_{AB}
$ and $
    {\cal Z}_{AB}(\rho)
    =
    F\rho F.
$
This is a random local phase channel with probability \(\cos^2\Theta\) one
applies \(I_A\otimes I_B\), and with probability \(\sin^2\Theta\) one applies
\(Z_A\otimes Z_B\). Hence \(\Lambda_*\in{\rm LP}\), and in particular it is
LOCC and PPT-Choi.

Its normalized Choi state is
\begin{equation}
    \omega_{\Lambda_*}
    =
    \cos^2\Theta\,
    |I\rangle\rangle\langle\langle I|
    +
    \sin^2\Theta\,
    |F\rangle\rangle\langle\langle F| .
\end{equation}
On the other hand,
\begin{align}
    \omega_{{\cal U}_\Theta}
   &=|U_\Theta\rangle\rangle\langle\langle U_\Theta|
    \nonumber\\
    &=
    \cos^2\Theta\,
    |I\rangle\rangle\langle\langle I|
    +
    \sin^2\Theta\,
    |F\rangle\rangle\langle\langle F|
    \nonumber\\
    &\quad
    -i\cos\Theta\sin\Theta\,
    |I\rangle\rangle\langle\langle F|
    +
    i\cos\Theta\sin\Theta\,
    |F\rangle\rangle\langle\langle I| .
\end{align}
Therefore
\begin{equation}
    \omega_{{\cal U}_\Theta}-\omega_{\Lambda_*}
    =
    -i\cos\Theta\sin\Theta\,
    |I\rangle\rangle\langle\langle F|
    +
    i\cos\Theta\sin\Theta\,
    |F\rangle\rangle\langle\langle I| .
\end{equation}
In the two-dimensional subspace spanned by
\(\{|I\rangle\rangle,|F\rangle\rangle\}\), this operator has eigenvalues
$
    \pm |\cos\Theta\sin\Theta| ,
$
and it vanishes on the orthogonal complement. Hence
\begin{equation}
    D_{\rm Choi}({\cal U}_\Theta,\Lambda_*)
    =
    \frac12
    \left\|
    \omega_{{\cal U}_\Theta}-\omega_{\Lambda_*}
    \right\|_1
    =
    |\cos\Theta\sin\Theta|
    =
    \frac{|\sin2\Theta|}{2}.
\end{equation}
Combining the PPT lower bound with this local-phase achievable channel yields
\begin{equation}
    {
    \inf_{\Lambda\in{\rm PPT\text{-}Choi}}
    D_{\rm Choi}({\cal U}_\Theta,\Lambda)
    =
    \inf_{\Lambda\in{\rm LP}}
    D_{\rm Choi}({\cal U}_\Theta,\Lambda)
    =
    \frac{|\sin2\Theta|}{2}
    } .
\end{equation}
This proves the noiseless Choi equivalence between the full PPT-Choi
relaxation and the path-local phase model.

\section{Two-mode path-transfer and gravitational-displacement estimates}
\label{app:path_transfer}
\subsection{Gaussian overlap and orthonormal path modes}
Let $d$ denote the arm separation, distinct from the intermass separation $L$ used in the main text. Consider normalized real Gaussians of rms width $\sigma_x$,
\begin{equation}
\psi_{0,1}(x)=\frac1{(2\pi\sigma_x^2)^{1/4}}
\exp\!\left[-\frac{(x\pm d/2)^2}{4\sigma_x^2}\right].
\label{eq:gaussian_path_modes}
\end{equation}
Their overlap is
$
S=\langle\psi_0|\psi_1\rangle=e^{-d^2/(8\sigma_x^2)}.
$
For $d\gg\sigma_x$ this overlap is small, but the pair is not exactly orthogonal. In particular, a bare $\langle\psi_1|V|\psi_0\rangle$ matrix element contains an unphysical overlap contribution even for $V=V_0I$. A more appropriate orthonormal two-mode basis is
\begin{equation}
|e_0\rangle=|\psi_0\rangle,\qquad
|e_1\rangle=\frac{|\psi_1\rangle-S|\psi_0\rangle}{\sqrt{1-S^2}}.
\label{eq:GS_modes}
\end{equation}
Let $V(x)$ denote the residual multiplication potential after removing any arbitrary common scalar potential, and suppose $|V(x)|\le V_*$ on the spatial region used for the estimate. For real positive Gaussians, the Cauchy/triangle estimate gives
\begin{equation}
|\langle\psi_1|V|\psi_0\rangle|
\le V_*\int dx\,\psi_1(x)\psi_0(x)=V_*S,
\end{equation}
and $|\langle\psi_0|V|\psi_0\rangle|\le V_*$. Hence
\begin{align}
|\langle e_1|V|e_0\rangle|
&=\frac{|\langle\psi_1|V|\psi_0\rangle-S\langle\psi_0|V|\psi_0\rangle|}{\sqrt{1-S^2}}
\nonumber\\
&\le\frac{2V_*S}{\sqrt{1-S^2}}.
\label{eq:app_GS_hopping_bound}
\end{align}
This conservative factor of two is retained in the plotted estimate. The bound is a two-mode matrix-element estimate under the indicated boundedness hypothesis, and it is not a claim that gravitational motion is exactly confined to these two modes.

\subsection{Perturbative transfer probability and limitations}
If the effective two-mode dynamics is well described by a prescribed background trajectory plus a small off-diagonal residual coupling, first-order time-dependent perturbation theory yields
\begin{equation}
P_{\rm hop}^{\rm(est)}\lesssim\left(\frac1\hbar\int_0^\tau\!dt\,|V_{10}(t)|\right)^2.
\label{eq:hop_integral_estimate}
\end{equation}
For a constant characteristic bound $V_*$ this reduces to Eq.~\eqref{eq:hop_prob_bound}. Additional free evolution, controls or leakage outside the two-mode subspace are not estimated by this expression and should be constrained by separate single-interferometer measurements. In particular, when the right side becomes of order unity, the perturbative interpretation must not be used, and the plotted parameter domain is restricted to where the estimate is small.

A classical Newtonian force can deflect the centers without implementing a projective hop. If the distance between masses is always at least $R_{\min}$ during the hold, the magnitude of their acceleration is at most $Gm/R_{\min}^2$ in the point-mass approximation. With an initially vanishing uncorrected gravitational velocity, the free-hold displacement obeys
\begin{equation}
\delta x\lesssim\frac{Gm\tau^2}{2R_{\min}^2}.
\label{eq:app_displacement_bound}
\end{equation}
It is useful to compare both $\delta x/d$ and $\delta x/\sigma_x$, since substantial displacement within one arm may affect contrast without causing occupation transfer to the other arm. Experimental controls may compensate accelerations or introduce other motion. Eq.~\eqref{eq:app_displacement_bound} is not a substitute for a complete trajectory simulation.

For illustrative held collinear parameters motivated by the BMV-scale ranges of Refs.~\cite{bose_spin_2017,boseRoadmap2025,boseWhitepaper2026}, take $m=10^{-14}\,\mathrm{kg}$, $L=40\,\mu\mathrm m$, $d=10\,\mu\mathrm m$ and $\tau=1\,\mathrm s$. Choosing the additional model parameter $\sigma_x=1\,\mu\mathrm m$ then gives $d/\sigma_x=10$, $S=e^{-12.5}\simeq3.73\times10^{-6}$, and $\Gamma_g\tau\simeq0.052741$. Taking $R_{\min}=30\,\mu\mathrm m$ and the deliberately generous potential scale $V_*=Gm^2/R_{\min}$ gives $V_*\tau/\hbar\simeq2.11$, yet the off-diagonal factor in Eq.~\eqref{eq:app_GS_hopping_bound} makes the perturbative transfer estimate of order $2.5\times10^{-10}$. Here $V_*$ is a characteristic energy scale, rather than
a global bound on the point-mass Newtonian potential
over the full Gaussian tails. A rigorous application
requires a bounded effective potential or a controlled
treatment of the tails. Small position deflections alone
also do not bound differential momentum kicks or phase
gradients within each wavepacket.
The same conservative acceleration estimate gives $\delta x\lesssim3.71\times10^{-16}\,\mathrm m$, or $\delta x/\sigma_x\lesssim3.71\times10^{-10}$. For an illustrative local visibility $r=0.1$ (not an observed BMV visibility), the passive phase gap is $5.2643\times10^{-3}$. These scales support the no-hop approximation under the stated idealized conditions. They do not directly bound every possible classical channel or every non-phase noise source.

\subsection{Continuity under independently bounded modelling error}
Suppose a physically admissible simulator family $\mathfrak C_{\rm phys}$ satisfies a separately justified output-closeness condition: for every $\mathcal C\in\mathfrak C_{\rm phys}$ there exists an LP channel $\mathcal C_{\rm LP}$ with
\begin{equation}
D\!\left(\Dd_r\circ\mathcal C(\rho_1),\Dd_r\circ\mathcal C_{\rm LP}(\rho_1)\right)\le\epsilon_{\rm model}.
\label{eq:model_closeness_assumption}
\end{equation}
The triangle inequality and Theorem~\ref{thm:LP_phase_gap} then give the conditional stability estimate
\begin{equation}
\inf_{\mathcal C\in\mathfrak C_{\rm phys}}D\!\left(\sigma_1,\Dd_r\circ\mathcal C(\rho_1)\right)
\ge\max\{0,g_{\rm LP}-\epsilon_{\rm model}\}.
\label{eq:model_stability_bound}
\end{equation}
Neither $P_{\rm hop}^{\rm(est)}$ nor $\delta x/\sigma_x$ alone implies Eq.~\eqref{eq:model_closeness_assumption}: it requires a controlled microscopic reduction or independent calibration. We include this result to make the additional assumption explicit, not as an unconditional extension of the LP theorem.

\subsection{A sufficient Hamiltonian bound for finite arm transfer}
For completeness, the output-closeness premise used in Sec.~\ref{sec:robustness_hopping} can be related to a controlled perturbative dynamics. Write the simulator generator as $\mathcal L_t=\mathcal L_{0,t}+\mathcal K_t$, where $\mathcal L_{0,t}$ generates a passive diagonal phase evolution interleaved with the same external dephasing and $\mathcal K_t(\rho)=-(i/\hbar)[H_{\rm off}(t),\rho]$. Let $\mathcal E_{t,s}$ denote the full CPTP propagator and $\mathcal E^0_{t,s}$ its reference counterpart. The variation-of-constants formula gives
\begin{equation}
(\mathcal E_{\tau,0}-\mathcal E^0_{\tau,0})(\rho)=\int_0^\tau dt\,\mathcal E_{\tau,t}\mathcal K_t\mathcal E^0_{t,0}(\rho).
\label{eq:duhamel_hop}
\end{equation}
For any density matrix $\omega$, $\|[H_{\rm off},\omega]\|_1\le2\|H_{\rm off}\|_\infty$. Trace-norm contraction of CPTP propagators on Hermitian operators therefore yields
\begin{equation}
D(\mathcal E_{\tau,0}(\rho),\mathcal E^0_{\tau,0}(\rho))\le\frac1\hbar\int_0^\tau dt\,\|H_{\rm off}(t)\|_\infty.
\label{eq:duhamel_hop_bound}
\end{equation}
Classical mixing preserves this bound whenever it holds uniformly (or in expectation) over the realized paths. Thus a physical passive model whose zero-hopping reference belongs to LP admits an explicit choice for the $\epsilon_C$ in Eq.~\eqref{eq:finite_hop_robustness_main}. The result does not follow solely from a measured population leakage probability and does not apply to active feedback models whose zero-hopping reference lies outside LP. The two-mode Gaussian estimate is one possible ingredient for controlling $\|H_{\rm off}\|_\infty$, provided the mode projection and other residual couplings are also bounded.

\section{Active no-hop feedback and the diagonal PPT relaxation}
\label{app:diag_ppt_proof}

In this appendix we analyze a broader no-hop class than the passive path-local phase channels used in the main analytic witness. The channels considered here are diagonal Schur multipliers constrained by PPT across the \(A|B\) cut. They include product-diagonal Kraus maps and active no-hop measurement-feedback phase operations. We derive an exact one-dimensional variational characterization, a global analytic lower bound, and the exact small-\(\Theta\) optimum.

\subsection{Diagonal no-hop maps as Schur multipliers}

We first make explicit how the no-hop condition is represented mathematically. Let
$x,y\in\{00,01,10,11\}$ label the four path states. A diagonal no-hop operation is a channel admitting Kraus operators diagonal in this basis,
$
K_\mu=\sum_x k_\mu(x)|x\rangle\langle x|.
$
Such a Kraus operator never maps one path state to another. Hence the channel preserves path populations and acts only by damping or rotating coherences.

Indeed, 
\begin{equation}
    \mathcal F_M(|x\rangle\langle y|)
=
M_{xy}|x\rangle\langle y|
\end{equation}
and 
$
M_{xy}
=
\sum_\mu k_\mu(x)k_\mu(y)^*.
$
Thus \(\mathcal F_M\) is a Schur multiplier in the path basis. Trace preservation gives \(M_{xx}=1\), so \(\mathcal F_M(|x\rangle\langle x|)=|x\rangle\langle x|\). Complete positivity is equivalent to \(M\succeq0\), since \(M\) is the Gram matrix of the vectors \(k(x)=(k_\mu(x))_\mu\). Therefore diagonal no-hop CPTP maps are exactly Schur multipliers by correlation matrices \(M\succeq0\) with unit diagonal.

To describe no-hop separable or feedback-type operations, we impose the stronger product-diagonal Kraus structure
\begin{equation}
K_\mu=A_\mu\otimes B_\mu,
\label{eq:product_diag_kraus}
\end{equation}
with
$
A_\mu=a_{\mu,0}|0\rangle\langle0|+a_{\mu,1}|1\rangle\langle1|
$ and $
B_\mu=b_{\mu,0}|0\rangle\langle0|+b_{\mu,1}|1\rangle\langle1|.
$
Then
\begin{equation}
M_{ab,cd}
=
\sum_\mu
a_{\mu,a}b_{\mu,b}
a_{\mu,c}^*b_{\mu,d}^*.
\label{eq:M_product_diag}
\end{equation}
This is the diagonal no-hop separable class.
In \(2\times2\), separability of this correlation matrix across the \(A|B\) label cut is equivalent to PPT. We therefore use the diagonal PPT relaxation
$
M\succeq0,$ $
M_{xx}=1,$ and $
M^{\Gamma_B}\succeq0,
$
where \((M^{\Gamma_B})_{ab,cd}=M_{ad,cb}\).

\subsection{Target and diagonal no-hop Schur channels}

Let \(\rho_1=|++\rangle\langle++|\) and \(U_\Theta=e^{i\Theta Z_AZ_B}\). In the computational basis ordered as \(|00\rangle,|01\rangle,|10\rangle,|11\rangle\), the dephased controlled-phase target is
\begin{equation}
\rho_{\rm tar}(\Theta,r)
=
\frac14
\begin{pmatrix}
1 & r e^{i2\Theta} & r e^{i2\Theta} & r^2\\
r e^{-i2\Theta} & 1 & r^2 & r e^{-i2\Theta}\\
r e^{-i2\Theta} & r^2 & 1 & r e^{-i2\Theta}\\
r^2 & r e^{i2\Theta} & r e^{i2\Theta} & 1
\end{pmatrix}.
\label{eq:diagPPT_target_matrix}
\end{equation}
Define
$
T=X\otimes I+I\otimes X,$ $
L=Y\otimes Z+Z\otimes Y,$ and  $
R=X\otimes X.
$
Then Eq.~\eqref{eq:diagPPT_target_matrix} can be written as
\begin{equation}
\rho_{\rm tar}(\Theta,r)
=
\frac14
\left[
I+r\cos(2\Theta)\,T
-r\sin(2\Theta)\,L
+r^2R
\right].
\label{eq:diagPPT_target_pauli}
\end{equation}

A diagonal no-hop Schur channel \(\mathcal F_M\) acts as
$
(\mathcal F_M(\rho))_{xy}=M_{xy}\rho_{xy},
$
and it commutes with the dephasing channel.
Here \(M\) is a correlation matrix satisfying \(M\succeq0\) and \(M_{xx}=1\).  Therefore, the simulated state induced by the classical model is
\begin{equation}
\rho_{\rm sim}(M,r)
=
\mathcal D_r(\mathcal F_M(\rho_1)).
\label{eq:diagPPT_sim_def}
\end{equation}

\subsection{Symmetry reduction}

The target state is invariant under the following operations: qubit swap, simultaneous bit flip $X_A X_B$, and the antiunitary map
\[
\mathcal A_A(O)=X_A O^* X_A ,
\]
where \(^*\) denotes complex conjugation in the computational basis. In the ordered basis \(|00\rangle,|01\rangle,|10\rangle,|11\rangle\), the last symmetry is \(O\mapsto P_AO^*P_A\), with \(P_A\) exchanging \(00\leftrightarrow10\) and \(01\leftrightarrow11\). It leaves \(T=XI+IX\), \(L=YZ+ZY\), and \(R=XX\) invariant, and hence leaves \(\rho_{\rm tar}\) invariant.
The diagonal PPT feasible set is invariant under the same transformations. Since the trace norm is convex and invariant under unitary and antiunitary conjugations, averaging any feasible \(M\) over these symmetries cannot increase the objective. Therefore, an optimizer may be taken in the form
\begin{equation}
M(\zeta,m)
=
\begin{pmatrix}
1&\zeta&\zeta&m\\
\zeta^*&1&m&\zeta^*\\
\zeta^*&m&1&\zeta^*\\
m&\zeta&\zeta&1
\end{pmatrix},
\qquad
\zeta=a+ib,\quad m\in\mathbb R.
\label{eq:symmetric_M}
\end{equation}
Equivalently,
\begin{equation}
M(\zeta,m)=I+aT-bL+mR.
\label{eq:symmetric_M_TLR}
\end{equation}

Direct diagonalization gives the spectra of \(M\) and its partial transpose as
\begin{equation}
\operatorname{spec}M
=
\{1-m,\,1-m,\,1+m\pm2|\zeta|\},
\label{eq:spec_M}
\end{equation}
and
\begin{equation}
\operatorname{spec}M^{\Gamma_B}
=
\{1+m\pm2a,\;1-m\pm2b\}.
\label{eq:spec_M_PPT}
\end{equation}
Setting
$
h=\frac{1-m}{2},
$
then \(0\le h\le1\), and the complete positive semidefinite/PPT conditions reduce to
\begin{equation}
|\zeta|\le1-h,
\qquad
|\operatorname{Im}\zeta|\le h.
\label{eq:zeta_feasible}
\end{equation}
Thus, for fixed \(h\), \(\zeta\) ranges over the disk of radius \(1-h\) truncated by the strip \(|\operatorname{Im}\zeta|\le h\).

For a symmetric feasible \(M\), the simulated state is
$
\rho_{\rm sim}
=
\frac14
\left[
I+raT-rbL+r^2mR
\right] 
$
and let
\begin{equation}
d=|e^{i2\Theta}-\zeta|.
\label{eq:d_def}
\end{equation}
The four eigenvalues of \(\rho_{\rm tar}-\rho_{\rm sim}\) are
\begin{equation}
\frac14
\left\{
-2r^2h,\,
-2r^2h,\,
2r^2h+2rd,\,
2r^2h-2rd
\right\}.
\label{eq:difference_eigenvalues}
\end{equation}
Because \(|\zeta|\le1-h\), one has \(d\ge h\), and hence \(2r^2h\le2rd\) for \(0\le r\le1\). Therefore
\begin{equation}
D(\rho_{\rm tar},\rho_{\rm sim})
=
\frac r2(rh+d).
\label{eq:distance_rh_d}
\end{equation}

\subsection{Exact one-dimensional reduction}

Let
$
S=|\sin2\Theta|
$ and $
C=|\cos2\Theta|.
$
For fixed \(h\), let \(d_\Theta(h)\) be the Euclidean distance from the point \((C,S)\) to the feasible set
\begin{equation}
K_h
=
\{(a,b):a^2+b^2\le(1-h)^2,\ |b|\le h\}.
\label{eq:Kh_def}
\end{equation}
Elementary projection geometry gives
\begin{equation}
d_\Theta(h)=
\begin{cases}
S-h,
&
0\le h\le S^2/2,
\\[1ex]
\sqrt{\left(C-\sqrt{1-2h}\right)^2+(S-h)^2},
&
S^2/2\le h\le S/(1+S),
\\[1ex]
h,
&
S/(1+S)\le h\le1.
\end{cases}
\label{eq:dt_piecewise}
\end{equation}
Combining Eqs.~\eqref{eq:distance_rh_d} and \eqref{eq:dt_piecewise}, the exact optimum over the diagonal PPT relaxation is
\begin{equation}
g_{\rm diagPPT}(\Theta,r)
=
\frac r2
\min_{0\le h\le1}
\left[
rh+d_\Theta(h)
\right].
\label{eq:diagPPT_exact_1D}
\end{equation}
This is an exact one-dimensional characterization valid for arbitrary \(\Theta\) and \(r\).

For \(0<r<1\) and \(0<S<1\), the objective decreases on the first interval and increases on the third. Hence the minimizer lies in the middle interval and satisfies
\begin{equation}
\frac{
1+S-h_*-\dfrac{C}{\sqrt{1-2h_*}}
}{
\sqrt{
\left(C-\sqrt{1-2h_*}\right)^2
+
(S-h_*)^2
}
}
=
r.
\label{eq:hstar_condition}
\end{equation}
The corresponding optimal parameters are
\begin{equation}
m_*=1-2h_*,
\qquad
\zeta_*
=
\operatorname{sgn}(\cos2\Theta)\sqrt{1-2h_*}
+
i\,\operatorname{sgn}(\sin2\Theta)\,h_*.
\label{eq:zeta_star}
\end{equation}

Two checks follow immediately. At \(\Theta=\pi/4\), one has \(h_*=1/2\), giving
$
g_{\rm diagPPT}(\pi/4,r)
=
\frac{r(1+r)}{4}.
$
For \(r=1\), the optimum reduces to
$
g_{\rm diagPPT}(\Theta,1)
=
\frac12|\sin2\Theta|,
$
matching the noiseless Choi optimum.

\subsection{Small-phase asymptotics}

Let \(\tau_\Theta=|2\Theta|\) and set \(h=k\tau_\Theta\). In the middle region of Eq.~\eqref{eq:dt_piecewise},
\begin{equation}
d_\Theta(k\tau_\Theta)
=
\tau_\Theta
\sqrt{k^2+(1-k)^2}
+
O(\tau_\Theta^2).
\label{eq:dt_small_theta}
\end{equation}
Thus
\begin{equation}
g_{\rm diagPPT}(\Theta,r)
=
\frac r2
\min_{0\le k\le1}
\left[
rk+\sqrt{k^2+(1-k)^2}
\right]
|2\Theta|
+
O(\Theta^2).
\label{eq:small_theta_min}
\end{equation}
Let
$
s_r=\sqrt{2-r^2}, 
$
and the minimizer for \(0<r<1\) is
$
k_*(r)
=
\frac12
\left(
1-\frac r{s_r}
\right).
$
At this point,
$
\sqrt{k_*^2+(1-k_*)^2}
=
\frac1{s_r}.
$
Therefore
\begin{equation}
g_{\rm diagPPT}(\Theta,r)
=
\frac r2
\left(
r+\sqrt{2-r^2}
\right)
|\Theta|
+
O(\Theta^2).
\label{eq:diagPPT_small_theta}
\end{equation}

\subsection{Analytic global dual lower bound with dephasing}

We now prove the global lower bound (with dephasing) in Eq.~\eqref{eq:diagPPT_global_bound}. Define
$
\alpha=\frac{s_r-r}{2}
$ and $
\beta=\frac{s_r+r}{2}.
$
Then \(\alpha^2+\beta^2=1\) and \(\beta-\alpha=r\). Choose $\epsilon=+1$ when $\sin(2\Theta)\geq0$ and
$\epsilon=-1$ otherwise. Define the witness operator
\begin{equation}
W_\epsilon
=
-\frac{I-R}{2}
+
\frac{s_r-r}{4}T
-
\epsilon
\frac{s_r+r}{4}L .
\label{eq:dual_witness_diagPPT}
\end{equation}
Equivalently,
\begin{equation}
W_\epsilon
=
\frac12
\begin{pmatrix}
-1&\alpha+i\epsilon\beta&\alpha+i\epsilon\beta&1\\
\alpha-i\epsilon\beta&-1&1&\alpha-i\epsilon\beta\\
\alpha-i\epsilon\beta&1&-1&\alpha-i\epsilon\beta\\
1&\alpha+i\epsilon\beta&\alpha+i\epsilon\beta&-1
\end{pmatrix}.
\label{eq:dual_witness_matrix}
\end{equation}
On the \(R=-1\) subspace, \(W_\epsilon=-I\). On the \(R=+1\) subspace, \(T/2\) and \(L/2\) act as anticommuting Pauli operators, while \(\alpha^2+\beta^2=1\). Hence
\begin{equation}
\operatorname{spec}W_\epsilon=\{1,-1,-1,-1\},
\qquad
\|W_\epsilon\|_\infty=1.
\label{eq:dual_witness_norm}
\end{equation}

For a symmetrized feasible matrix \(M(\zeta,m)\), with \(\zeta=a+ib\) and \(m=1-2h\), feasibility gives \(a\le1-h\) and \(\epsilon b\le h\). A direct calculation yields
\begin{align}
\frac12
\operatorname{Tr}
[W_\epsilon
(\rho_{\rm tar}-\rho_{\rm sim})]
={}&
\frac r2
\left[
\alpha(\cos2\Theta-1)
+
\beta|\sin2\Theta|
\right]
\nonumber\\
&\quad +\frac r2
\left[
\alpha(1-h-a)
+
\beta(h-\epsilon b)
\right].
\label{eq:witness_eval_diagPPT}
\end{align}
Here the second term is nonnegative. Since the witness is invariant under the same symmetries used in the reduction, the bound holds for every feasible \(M\). Therefore trace-distance duality gives
\begin{equation}
g_{\rm diagPPT}(\Theta,r)
\ge
\frac r4
\left[
(s_r+r)|\sin2\Theta|
-
(s_r-r)(1-\cos2\Theta)
\right].
\label{eq:diagPPT_global_bound_app}
\end{equation}
Near \(\Theta=0\), this becomes
$
g_{\rm diagPPT}(\Theta,r)
\ge
\frac{r(s_r+r)}2|\Theta|
-
\frac{r(s_r-r)}2\Theta^2
+
O(|\Theta|^3),
$
which matches the feasible construction at linear order.

\subsection{Exact LOCC achievability and feedback interpretation}
\label{app:diagPPT_achievability_feedback}

We now show that the diagonal-PPT optimum admits an
explicit finite-round LOCC implementation. The feedback
construction realizes the relevant boundary Schur matrices
exactly, rather than only to leading order in the phase.
Consequently, the optimal errors over diagonal LOCC,
diagonal SEP, and diagonal PPT channels coincide for the
terminal-channel optimization considered here.

Define
$
C=|\cos(2\Theta)|
$, $
S=|\sin(2\Theta)|,
$
and choose signs
\begin{equation}
\eta=
\begin{cases}
+1,&\cos(2\Theta)\geq0,\\
-1,&\cos(2\Theta)<0,
\end{cases}
\qquad
\epsilon=
\begin{cases}
+1,&\sin(2\Theta)\geq0,\\
-1,&\sin(2\Theta)<0.
\end{cases}
\end{equation}
Thus $\cos(2\Theta)=\eta C$ and
$\sin(2\Theta)=\epsilon S$.

For $0\leq h\leq1/2$, set
\begin{equation}
m_h=1-2h,
\qquad
\zeta_h=\eta\sqrt{1-2h}+i\epsilon h,
\label{eq:achievable_params_feedback}
\end{equation}
and consider the Schur matrix
\begin{equation}
M_h=
\begin{pmatrix}
1&\zeta_h&\zeta_h&1-2h\\
\zeta_h^*&1&1-2h&\zeta_h^*\\
\zeta_h^*&1-2h&1&\zeta_h^*\\
1-2h&\zeta_h&\zeta_h&1
\end{pmatrix}.
\label{eq:achievable_Mh_feedback}
\end{equation}
Its parameters satisfy
\begin{equation}
|\zeta_h|=1-h,
\qquad
|\operatorname{Im}\zeta_h|=h.
\end{equation}
In particular,
\begin{equation}
\operatorname{spec}M_h
=
\{0,\,2h,\,2h,\,4(1-h)\},
\label{eq:spec_Mh}
\end{equation}
and, independently of the choices of $\eta$ and $\epsilon$,
\begin{equation}
\operatorname{spec}M_h^{\Gamma_B}
=
\left\{
0,\,
4h,\,
(1+\sqrt{1-2h})^2,\,
(1-\sqrt{1-2h})^2
\right\}.
\label{eq:spec_Mh_PPT}
\end{equation}
Hence $M_h$ is a feasible diagonal PPT correlation matrix.
Since $M_h/4$ is a two-qubit state, PPT is equivalent to
separability across its path-label bipartition. Below we
establish the stronger, constructive statement that the
associated channel is LOCC.

\paragraph{Exact measurement-and-feedback realization.}

Define local path-measurement operators
\begin{equation}
D_s(\mu)
=
\operatorname{diag}
\left(
\sqrt{\frac{1+s\mu}{2}},
\sqrt{\frac{1-s\mu}{2}}
\right),
\qquad s=\pm1,
\qquad 0\leq\mu\leq1,
\label{eq:weak_measurement_Ds}
\end{equation}
and conditional phase rotations
\begin{equation}
V_s(\phi)=e^{is\phi Z/2}.
\label{eq:feedback_phase_Vs}
\end{equation}
Using a shared random bit, choose with equal probability
which party performs the measurement. That party sends its
outcome $s$ to the other party, which applies $V_s(\phi)$.
The resulting channel has Kraus operators
\begin{equation}
L_{A,s}
=
\frac{1}{\sqrt2}D_s(\mu)\otimes V_s(\phi),
\qquad
L_{B,s}
=
\frac{1}{\sqrt2}V_s(\phi)\otimes D_s(\mu).
\label{eq:weak_feedback_kraus}
\end{equation}
They obey
\begin{equation}
\sum_{s=\pm1}
\left(
L_{A,s}^{\dagger}L_{A,s}
+
L_{B,s}^{\dagger}L_{B,s}
\right)=I.
\end{equation}
This is a finite-round LOCC protocol, and every Kraus
operator is product diagonal. Measurement back-action is
included explicitly through $D_s(\mu)$.

Direct evaluation gives the symmetric Schur parameters
\begin{align}
\zeta_{\rm wf}
&=
\frac12\left(\sqrt{1-\mu^2}+\cos\phi\right)
+\frac{i}{2}\mu\sin\phi,
\label{eq:zeta_weak_feedback}\\
m_{\rm wf}
&=
\sqrt{1-\mu^2}\cos\phi.
\label{eq:m_weak_feedback}
\end{align}
Choose
\begin{equation}
\mu=\sqrt{2h},
\qquad
\phi=\eta\epsilon\,\arcsin\sqrt{2h}.
\label{eq:weak_feedback_choice}
\end{equation}
For every $0\leq h\leq1/2$, this yields exactly
\begin{equation}
\zeta_{\rm wf}
=
\sqrt{1-2h}+i\eta\epsilon h,
\qquad
m_{\rm wf}=1-2h.
\end{equation}

Finally, apply the deterministic local unitary
\begin{equation}
P_\eta=
\begin{cases}
I_A\otimes I_B,&\eta=+1,\\
Z_A\otimes Z_B,&\eta=-1.
\end{cases}
\end{equation}
For $\eta=-1$, this changes the sign of every single-bit
coherence while leaving the two-bit coherences unchanged.
Thus the final Kraus operators
\begin{equation}
\widetilde L_{A,s}=P_\eta L_{A,s},
\qquad
\widetilde L_{B,s}=P_\eta L_{B,s}
\end{equation}
generate precisely the matrix $M_h$ in
Eq.~\eqref{eq:achievable_Mh_feedback}.
The realization is exact throughout $0\leq h\leq1/2$.

\paragraph{Attainment of the exact optimum.}

For the family above,
Eq.~\eqref{eq:distance_rh_d} gives
\begin{align}
D\bigl(\rho_{\rm tar},\rho_{\rm sim}(M_h,r)\bigr)
&=
\frac r2\left[rh+|e^{i2\Theta}-\zeta_h|\right]
\nonumber\\
&=
\frac r2
\left[
rh+
\sqrt{
\left(C-\sqrt{1-2h}\right)^2+(S-h)^2
}
\right].
\label{eq:Mh_distance_exact}
\end{align}

The preceding projection analysis shows that an optimizer
of the full diagonal-PPT problem can be chosen with
\begin{equation}
h_*\in
\left[\frac{S^2}{2},\frac{S}{1+S}\right]
\subseteq[0,1/2].
\end{equation}
On this interval, the nearest feasible point is precisely
the boundary point used in
Eq.~\eqref{eq:achievable_params_feedback}.
For $0<r<1$ and $0<S<1$, the minimizer lies in the
interior.
Consequently,
\begin{equation}
g_{\rm diagPPT}(\Theta,r)
=
\frac r2
\min_{\frac{S^2}{2}\leq h\leq\frac{S}{1+S}}
\left[
rh+
\sqrt{
\left(C-\sqrt{1-2h}\right)^2+(S-h)^2
}
\right].
\label{eq:exact_feedback_optimum}
\end{equation}
The LOCC construction above attains this value by choosing
$h=h_*$.

The boundary cases are also covered. If $S=0$, take
$h_*=0$, obtaining zero error. If $S=1$, take $h_*=1/2$,
which gives $g_{\rm diagPPT}=r(1+r)/4$.
For $r=1$, choosing $h_*=S^2/2$ gives
$g_{\rm diagPPT}=S/2$. For $r=0$, every simulated output
and the target equal $I/4$.

Let $g_{\rm diagLOCC}$ and $g_{\rm diagSEP}$ denote the
same objective with the corresponding restrictions on
the simulator. The inclusions of the feasible classes give
\begin{equation}
g_{\rm diagPPT}
\leq g_{\rm diagSEP}
\leq g_{\rm diagLOCC}.
\end{equation}
Since the diagonal-PPT optimum has just been attained by
a diagonal LOCC protocol,
\begin{equation}
g_{\rm diagLOCC}(\Theta,r)
=
g_{\rm diagSEP}(\Theta,r)
=
g_{\rm diagPPT}(\Theta,r).
\label{eq:diagLOCC_SEP_PPT_equal_optima}
\end{equation}
This is an equality of optimal values for this target and
objective; it does not assert equality of all three
channel classes. Nor does terminal-channel achievability
by itself imply attainability under an additional,
prescribed continuous-time control-strength bound.

\paragraph{Weak-phase limit and physical interpretation.}

For fixed $0<r\leq1$, define
\begin{equation}
s_r=\sqrt{2-r^2},
\qquad
k_*(r)=\frac12\left(1-\frac r{s_r}\right).
\end{equation}
To obtain an asymptotically optimal explicit protocol,
choose
\begin{equation}
h=k_*(r)|2\Theta|
\end{equation}
for sufficiently small $|\Theta|$.
Near $\Theta=0$, one has $\eta=+1$, and the exact
feedback construction yields
\begin{equation}
\zeta_h
=
1-h+i\operatorname{sgn}(\Theta)h+O(h^2),
\qquad
m_h=1-2h.
\end{equation}
Writing $t=|2\Theta|$, substitution into
Eq.~\eqref{eq:Mh_distance_exact} gives
\begin{equation}
D\bigl(\rho_{\rm tar},\rho_{\rm sim}(M_h,r)\bigr)
=
\frac r2
\left[
rk_*(r)
+
\sqrt{k_*^2(r)+(1-k_*(r))^2}
\right]t
+
O(t^2).
\end{equation}
The coefficient in brackets equals $(r+s_r)/2$.
Together with the dual lower bound, this proves
\begin{equation}
g_{\rm diagLOCC}(\Theta,r)
=
g_{\rm diagSEP}(\Theta,r)
=
g_{\rm diagPPT}(\Theta,r)
=
\frac r2
\left(r+\sqrt{2-r^2}\right)|\Theta|
+
O(\Theta^2).
\label{eq:Mh_distance_small_theta}
\end{equation}

The feedback mechanism also makes the disturbance cost
explicit. For small $h$,
\begin{equation}
\mu=\sqrt{2h},
\qquad
\phi=\operatorname{sgn}(\Theta)
\left[\sqrt{2h}+O(h^{3/2})\right].
\end{equation}
The conditional phase response is proportional to
$\mu\sin\phi$, whereas measurement back-action and the
random feedback phases produce coherence losses of
order $\mu^2$ and $\phi^2$, respectively.
For fixed $0<r<1$, the asymptotically optimal choice has
$h=O(|\Theta|)$, so both the conditional response and its
disturbance are linear in $|\Theta|$.
Active feedback reduces the leading coefficient relative
to passive random phases but cannot change the error to
$O(\Theta^2)$.

\paragraph{Scope of the fixed-environment comparison.}

For this Schur model, PPT of the output from $|++\rangle$
is also equivalent to separability of the Choi state of
the total noisy channel. In particular, in the all-time
regime $\gamma\geq\Gamma_g$, the total master equation
admits a continuous local-measurement-and-classical-feedback
realization if part of the observed dephasing is assigned
to measurement back-action.

Our results concern the stricter comparison in which the
full independently characterized external dephasing is
retained in addition to any mediator-induced noise.
The construction above includes the mediator's
measurement disturbance and is followed by that same
external dephasing. The positive simulation gap therefore
does not, without the environmental assumption, certify
nonseparability of the total noisy channel.

\section{General local which-path noise beyond symmetric Markovian dephasing}
\label{app:general_noise}

In the main text we use the symmetric Markovian dephasing model $r=e^{-2\gamma\tau}$ because it gives a transparent open-system dynamics and a closed analytic separability threshold. The channel-witness argument, however, does not rely on the exponential Markovian form. It only requires a weaker and physically natural condition for BMV interferometry: the external environment acts locally in the path basis, preserves path populations, and attenuates path coherences by an independently calibrated visibility.

\subsection{Local phase-covariant which-path noise}

Consider a general local phase-covariant noise channel $\mathcal N=\mathcal N_A\otimes\mathcal N_B$ acting after the gravitational or classical mediator dynamics. On each path qubit we assume
\begin{align}
\mathcal N_j(|0\rangle\langle0|)
&=
|0\rangle\langle0|,
&
\mathcal N_j(|1\rangle\langle1|)
&=
|1\rangle\langle1|,
\nonumber\\
\mathcal N_j(|0\rangle\langle1|)
&=
\kappa_j(\tau)|0\rangle\langle1|,
&
\mathcal N_j(|1\rangle\langle0|)
&=
\kappa_j^*(\tau)|1\rangle\langle0|.
\label{eq:general_local_dephasing}
\end{align}
Here $|\kappa_j(\tau)|\le1$. The Markovian model used in the main text is the special case $\kappa_A(\tau)=\kappa_B(\tau)=r=e^{-2\gamma\tau}$. More generally, $\kappa_j(\tau)$ may be non-exponential, non-Markovian, or complex. Its modulus is the calibrated single-particle visibility, while its phase is a local phase drift that can also be calibrated.

For the coherent input $\rho_1=|++\rangle\langle++|$, the quantum target is $\sigma_1=\mathcal N\circ\mathcal U_\Theta(\rho_1)$. The two $B$-coherences conditioned on the value of $A$ are
\begin{equation}
(\sigma_1)_{00,01}
=
\frac{\kappa_B}{4}e^{+i2\Theta},
\qquad
(\sigma_1)_{10,11}
=
\frac{\kappa_B}{4}e^{-i2\Theta}.
\label{eq:general_target_B_coherences}
\end{equation}
By contrast, for any path-local phase mediator $\mathcal C_{\rm LP}$, the simulated output $\tilde\sigma_1=\mathcal N\circ\mathcal C_{\rm LP}(\rho_1)$ satisfies
\begin{equation}
(\tilde\sigma_1)_{00,01}
=
(\tilde\sigma_1)_{10,11}.
\label{eq:general_LP_identity}
\end{equation}
This remains true because the same local $B$-coherence factor $\kappa_B$ multiplies both matrix elements.

Let $\kappa_B=|\kappa_B|e^{i\varphi_B}$. Define the phase-adapted witness
\begin{align}
W_B(\varphi_B)
&=
i e^{i\varphi_B}|00\rangle\langle01|
-
i e^{-i\varphi_B}|01\rangle\langle00|
\nonumber\\
&\quad
-
i e^{i\varphi_B}|10\rangle\langle11|
+
i e^{-i\varphi_B}|11\rangle\langle10|.
\label{eq:phase_adapted_WB}
\end{align}
It is Hermitian, block diagonal, and obeys $\|W_B(\varphi_B)\|_\infty=1$. For any density matrix $\rho$,
\begin{equation}
\operatorname{Tr}[W_B(\varphi_B)\rho]
=
2\,\operatorname{Im}\!\left(e^{-i\varphi_B}\rho_{00,01}\right)
-
2\,\operatorname{Im}\!\left(e^{-i\varphi_B}\rho_{10,11}\right).
\label{eq:general_WB_expectation}
\end{equation}
Using Eq.~\eqref{eq:general_target_B_coherences}, one obtains $\operatorname{Tr}[W_B(\varphi_B)\sigma_1]=|\kappa_B|\sin2\Theta$. Using Eq.~\eqref{eq:general_LP_identity}, one obtains $\operatorname{Tr}[W_B(\varphi_B)\tilde\sigma_1]=0$ for every path-local phase mediator. Hence
\begin{equation}
D(\sigma_1,\tilde\sigma_1)
\ge
\frac12
\left|
\operatorname{Tr}[W_B(\varphi_B)(\sigma_1-\tilde\sigma_1)]
\right|
=
\frac{|\kappa_B|}{2}|\sin2\Theta|.
\label{eq:general_kappa_gap_B}
\end{equation}
Taking the infimum over all path-local phase mediators gives
\begin{equation}
{
g_{\rm LP}^{(B)}(\tau)
\ge
\frac{|\kappa_B(\tau)|}{2}|\sin2\Theta|.
}
\label{eq:general_kappa_gap_main}
\end{equation}

Similarly, using the $A$-coherence witness
\begin{align}
W_A(\varphi_A)
&=
i e^{i\varphi_A}|00\rangle\langle10|
-
i e^{-i\varphi_A}|10\rangle\langle00|
\nonumber\\
&\quad
-
i e^{i\varphi_A}|01\rangle\langle11|
+
i e^{-i\varphi_A}|11\rangle\langle01|,
\label{eq:phase_adapted_WA}
\end{align}
with $\kappa_A=|\kappa_A|e^{i\varphi_A}$, one obtains
\begin{equation}
g_{\rm LP}^{(A)}(\tau)
\ge
\frac{|\kappa_A(\tau)|}{2}|\sin2\Theta|.
\label{eq:general_kappa_gap_A}
\end{equation}
Thus, for asymmetric local dephasing, the experimentally stronger witness may be chosen:
\begin{equation}
g_{\rm LP}(\tau)
\ge
\frac{\max\{|\kappa_A(\tau)|,|\kappa_B(\tau)|\}}{2}
|\sin2\Theta|.
\label{eq:asymmetric_best_gap}
\end{equation}
The symmetric Markovian result in the main text is recovered by setting $\kappa_A=\kappa_B=r=e^{-2\gamma\tau}$.

Equation~\eqref{eq:general_kappa_gap_main} shows that the witness depends only on the calibrated single-particle coherence visibility. The decay need not be Markovian. If a local environment produces $|0\rangle\langle1|\mapsto \kappa_j(\tau)|0\rangle\langle1|$ with an arbitrary experimentally measured function $\kappa_j(\tau)$, then the same proof applies.

Thus the Markovian formula $g_{\rm LP}\ge \frac12 e^{-2\gamma\tau}|\sin2\Theta|$ is a convenient representative of the more general expression
\begin{equation}
g_{\rm LP}(\tau)
\ge
\frac{|\kappa(\tau)|}{2}|\sin2\Theta|.
\label{eq:nonmarkovian_visibility_gap}
\end{equation}
The essential physical requirement is not exponential decay, but local which-path decoherence with known visibility.

\subsection{Local correlated phase noise}

The result also tolerates shared classical fluctuations provided they are local additive phase fluctuations. For example, an external environment may apply random local phases $U_{\rm env}(\lambda)=U_A^{\rm env}(\lambda)\otimes U_B^{\rm env}(\lambda)$, where the same random variable $\lambda$ may be shared between the two sides. Such noise can generate correlated local phase fluctuations, but it does not produce a nonlocal conditional phase.

In particular, the coherences $|00\rangle\langle01|$ and $|10\rangle\langle11|$ acquire the same local $B$-phase factor. Therefore the path-local identity $(\tilde\sigma_1)_{00,01}=(\tilde\sigma_1)_{10,11}$ is preserved, and the witness argument goes through with the appropriate calibrated visibility.

This distinction is important. Correlated local dephasing is still compatible with the classical path-local benchmark. A genuinely nonlocal environmental phase proportional to $Z_AZ_B$, by contrast, has the same structure as the gravitational signal and must be independently bounded as a systematic effect.

\subsection{Population-changing noise and path hopping}

The analytic witness assumes that the external environment preserves the path populations. This is the natural effective noise model for BMV interferometry: dominant environmental mechanisms acquire which-path information and suppress off-diagonal path coherences, while leaving the occupation of each interferometer arm essentially unchanged.

If the environment or a proposed simulator induces path hopping or population transfer, additional diagnostics are required. The path-classical input
\begin{equation}
\rho_2
=
\frac12
\left(
|00\rangle\langle00|
+
|11\rangle\langle11|
\right)
\label{eq:general_noise_rho2}
\end{equation}
is useful for this purpose. The ideal gravitational channel and every path-local phase channel preserve $\rho_2$. If a simulated output $\tilde\sigma_2$ has leakage into the odd-parity subspace $\Pi_{\rm leak}=|01\rangle\langle01|+|10\rangle\langle10|$, with $p_{\rm leak}=\operatorname{Tr}[\Pi_{\rm leak}\tilde\sigma_2]$, then
\begin{equation}
D(\rho_2,\tilde\sigma_2)
\ge
p_{\rm leak}.
\label{eq:general_leakage_bound}
\end{equation}
This diagnostic detects final leakage into the odd-parity subspace. It does not detect every path-changing process: parity-preserving pair flips and excursions followed by return can evade this test. Excluding arbitrary hopping strategies requires additional dynamical assumptions or diagnostics.

Small deviations from ideal phase-covariant noise can also be bounded perturbatively. Let $\mathcal N_{\rm pc}$ be the calibrated population-preserving phase-covariant part and let $\mathcal N_{\rm exp}$ be the actual environmental channel. Define $\epsilon_{\rm env}=\frac12\|\mathcal N_{\rm exp}-\mathcal N_{\rm pc}\|_\diamond$. Then, for any fixed path-local phase mediator $\mathcal C_{\rm LP}$, the triangle inequality gives
\begin{align}
&
D\!\left(
\mathcal N_{\rm exp}\circ\mathcal U_\Theta(\rho_1),
\mathcal N_{\rm exp}\circ\mathcal C_{\rm LP}(\rho_1)
\right)
\nonumber\\
&\quad\ge
D\!\left(
\mathcal N_{\rm pc}\circ\mathcal U_\Theta(\rho_1),
\mathcal N_{\rm pc}\circ\mathcal C_{\rm LP}(\rho_1)
\right)
-
2\epsilon_{\rm env}.
\label{eq:triangle_env_error}
\end{align}
Therefore,
\begin{equation}
g_{\rm LP}^{\rm exp}
\ge
\frac{|\kappa_B|}{2}|\sin2\Theta|
-
2\epsilon_{\rm env},
\label{eq:robust_gap_env}
\end{equation}
or similarly with $\kappa_A$ using the $A$-coherence witness. Hence small non-phase-covariant imperfections reduce the certified gap continuously rather than invalidating the argument.

A qualitatively different situation occurs if the external environment itself generates an uncalibrated nonlocal phase, for example a stochastic or coherent term proportional to $Z_AZ_B$. Such a process produces the same type of conditional phase pattern as the gravitational interaction. It is therefore not distinguishable from the gravitational signal by the coherence witness alone.

This is not a peculiarity of the present proposal. Any BMV phase experiment must independently bound non-gravitational interactions that generate a $Z_AZ_B$-type conditional phase. In the present framework, such terms should either be included in the calibrated background model or treated as systematic errors. The theorem in the main text assumes that the external environment is local in the path basis, while the candidate nonlocal conditional phase is the gravitational mediation being tested.

For BMV interferometry, local path dephasing is the natural effective model of dominant environmental noise. Gas scattering, blackbody radiation, uncontrolled which-path information and local phase fluctuations primarily suppress coherences between the two arms of each mass. They do not normally transfer a mesoscopic wavepacket from one interferometer arm to the other. In the path-qubit subspace, their leading effect is therefore $|0\rangle\langle1|\mapsto\kappa(\tau)|0\rangle\langle1|$, with path populations approximately preserved.

The Markovian dephasing model used in the main text is thus not essential to the witness, and it is the simplest analytically solvable representative of a broader class of local which-path noise. The operational requirement is that the corresponding visibility and any local phase drift are independently characterized, so that the gravitational mediator is tested separately from the environmental background.

\end{document}